\documentclass[sigconf,natbib=true]{acmart}

\usepackage{amsmath}
\usepackage{amssymb}
\usepackage{amsthm}
\usepackage{array}
\usepackage{booktabs}
\usepackage{tabularx}
\usepackage{multirow}
\usepackage[inline]{enumitem}
\usepackage{graphicx}
\usepackage{subcaption}
\usepackage{hyperref}
\usepackage{xcolor}
\usepackage{colortbl}
\usepackage{makecell}
\usepackage{tikz}
\usetikzlibrary{shapes.geometric, arrows.meta, positioning, calc, tikzmark}
\usepackage{pifont}
\usepackage{algorithm}
\usepackage{algpseudocode}

\graphicspath{{figures/}}

\renewcommand\footnotetextcopyrightpermission[1]{}
\newtheorem{definition}{Definition}

\newcommand{\Dpool}{\mathcal{D}}
\newcommand{\Tset}{\mathcal{T}}
\newcommand{\Cset}{\mathcal{C}}
\newcommand{\Rset}{\mathcal{R}}
\newcommand{\Aset}{\mathcal{A}}
\newcommand{\score}{s}
\newcommand{\prob}{p}
\newcommand{\acc}{\mathrm{Acc}}
\newcommand{\BER}{\mathrm{BER}}

\newcommand{\tllm}{t_{\mathrm{LLM}}}
\newcommand{\Tpx}{T_{\mathrm{proxy}}}

\usepackage{xspace}

\providecommand{\MethodName}[1]{#1}
\providecommand{\DatasetName}[1]{#1}
\providecommand{\ModelName}[1]{#1}

\newcommand{\methCSV}{\MethodName{CSV}\xspace}
\newcommand{\methBARGAIN}{\MethodName{BARGAIN}\xspace}
\newcommand{\methScaleDoc}{\MethodName{ScaleDoc}\xspace}
\newcommand{\methDefault}{\MethodName{ScaleDoc}\xspace}
\newcommand{\methBERCP}{\MethodName{Phase-2}\xspace}
\newcommand{\methTwoPhase}{\MethodName{Two-Phase}\xspace}
\newcommand{\methBERLB}{\MethodName{BER-LB}\xspace}
\newcommand{\methSUPG}{\MethodName{SUPG}\xspace}
\newcommand{\methLOTUS}{\MethodName{LOTUS}\xspace}
\newcommand{\methPalimpzest}{\MethodName{Palimpzest}\xspace}
\newcommand{\methDocETL}{\MethodName{DocETL}\xspace}
\newcommand{\methPLOP}{\MethodName{PLOP}\xspace}

\newcommand{\dsPubMed}{\DatasetName{PubMed}\xspace}
\newcommand{\dsGovReport}{\DatasetName{GovReport}\xspace}
\newcommand{\dsBigPatent}{\DatasetName{BigPatent}\xspace}

\newcommand{\oracleLLM}{\ModelName{Llama-3.1-70B-Instruct}\xspace}
\newcommand{\proxyLLM}{\ModelName{Llama-3.1-8B-Instruct}\xspace}
\newcommand{\NVEmbed}{\ModelName{NV-Embed}\xspace}
\newcommand{\CrossEncoder}{\ModelName{Cross-encoder}\xspace}
\newcommand{\ColBERT}{\ModelName{ColBERT}\xspace}

\newcommand{\CE}{\ModelName{CE}\xspace}      % cross-encoder, short
\newcommand{\CB}{\ModelName{CB}\xspace}      % ColBERT projection, short

\usepackage{etoolbox}

\algnewcommand{\algorithmicinput}{\textbf{Input:}}
\algnewcommand{\algorithmicoutput}{\textbf{Output:}}
\algnewcommand{\Input}{\item[\algorithmicinput]}
\algnewcommand{\Output}{\item[\algorithmicoutput]}

\makeatletter
\newcommand{\qmapdef}[3]{\expandafter\def\csname qmap@#1@#2\endcsname{#3}}
\newcommand{\qpubmed}[1]{Q\csname qmap@pubmed@#1\endcsname}
\newcommand{\qgov}[1]{Q\csname qmap@gov@#1\endcsname}
\newcommand{\qbp}[1]{Q\csname qmap@bp@#1\endcsname}

\qmapdef{pubmed}{3}{1}    \qmapdef{pubmed}{4}{2}
\qmapdef{pubmed}{6}{3}    \qmapdef{pubmed}{8}{4}
\qmapdef{pubmed}{9}{5}    \qmapdef{pubmed}{12}{6}
\qmapdef{pubmed}{13}{7}   \qmapdef{pubmed}{15}{8}
\qmapdef{pubmed}{17}{9}   \qmapdef{pubmed}{18}{10}
\qmapdef{pubmed}{22}{11}  \qmapdef{pubmed}{42}{12}
\qmapdef{pubmed}{44}{13}  \qmapdef{pubmed}{48}{14}
\qmapdef{pubmed}{53}{15}  \qmapdef{pubmed}{59}{16}
\qmapdef{pubmed}{64}{17}  \qmapdef{pubmed}{73}{18}
\qmapdef{pubmed}{78}{19}  \qmapdef{pubmed}{81}{20}

\qmapdef{gov}{0}{1}       \qmapdef{gov}{1}{2}
\qmapdef{gov}{2}{3}       \qmapdef{gov}{3}{4}
\qmapdef{gov}{4}{5}       \qmapdef{gov}{7}{6}
\qmapdef{gov}{11}{7}      \qmapdef{gov}{16}{8}
\qmapdef{gov}{18}{9}      \qmapdef{gov}{25}{10}
\qmapdef{gov}{30}{11}     \qmapdef{gov}{32}{12}
\qmapdef{gov}{34}{13}     \qmapdef{gov}{36}{14}
\qmapdef{gov}{42}{15}     \qmapdef{gov}{45}{16}
\qmapdef{gov}{48}{17}     \qmapdef{gov}{50}{18}
\qmapdef{gov}{58}{19}     \qmapdef{gov}{68}{20}

\qmapdef{bp}{0}{1}        \qmapdef{bp}{1}{2}
\qmapdef{bp}{2}{3}        \qmapdef{bp}{3}{4}
\qmapdef{bp}{4}{5}        \qmapdef{bp}{9}{6}
\qmapdef{bp}{11}{7}       \qmapdef{bp}{13}{8}
\qmapdef{bp}{15}{9}       \qmapdef{bp}{20}{10}
\qmapdef{bp}{30}{11}      \qmapdef{bp}{35}{12}
\qmapdef{bp}{38}{13}      \qmapdef{bp}{41}{14}
\qmapdef{bp}{46}{15}      \qmapdef{bp}{51}{16}
\qmapdef{bp}{55}{17}      \qmapdef{bp}{59}{18}
\qmapdef{bp}{61}{19}      \qmapdef{bp}{65}{20}
\makeatother

\title{Fast LLM-Based Semantic Filtering: \\ From a Unified Framework to an Adaptive Two-Phase Method}

\author{Kyoungmin Kim}
\affiliation{\institution{EPFL}\city{Lausanne}\country{Switzerland}}
\email{kyoung-min.kim@epfl.ch}

\author{Martin Catheland}
\affiliation{\institution{EPFL}\city{Lausanne}\country{Switzerland}}
\email{martin.catheland@epfl.ch}

\author{Anastasia Ailamaki}
\affiliation{\institution{EPFL}\city{Lausanne}\country{Switzerland}}
\email{anastasia.ailamaki@epfl.ch}

\begin{abstract}
Evaluating a natural-language yes/no predicate over a document corpus
under a corpus-level accuracy target --- the \emph{semantic filter}
--- is a cornerstone of LLM-based data processing. 
The naive approach is to call the Large Language Model (LLM) on every document, which is considered as the oracle but prohibitively expensive as an online processing. Fast-proxy-then-LLM \emph{cascades} are the standard remedy but, as deployed
today, leave four structural limitations on the table. \emph{(1)}~Each
existing family --- model-free clustering, prebuilt small-LLM proxies,
and online-trained proxies --- commits to a single representation
and a fixed pipeline, and so wins on only a narrow query regime.
\emph{(2)}~The strongest online proxy invests heavily in
a custom-designed training scheme while leaving established
representation-learning tools on the table; its bi-encoder over
dense embeddings captures only topical similarity, missing the
token-level evidence and richer semantics many predicates require.
\emph{(3)}~The same proxy is trained against binary yes/no labels, weighting every
document identically, and wasting the LLM's per-document confidence at
exactly the boundary documents the proxy most needs to learn.
{\emph{(4)}~Existing
calibrations are too conservative. Proxy scores are aligned to the oracle LLM and a threshold is chosen to cascade lower-score documents to the oracle to meet accuracy target, which is often too larger than necessary, inflating the cascade cost. 
We observe a common root cause: they do not separate when the
proxy is genuinely uncertain on those documents from when the labeled
sample is simply too small to be sure, and treat both with the
same safety margin.}

{We address these limitations as follows.
\emph{(1)}~We adapt to each query's difficulty by composing
families: we first try lightweight model-free clustering, and escalate
to an online-trained proxy only when clustering becomes less efficient to classify all documents confidently. 
We share oracle LLM calls across phases to reduce the transition cost.
\emph{(2)}~For that online proxy, we replace
the cosine bi-encoder with a hybrid of off-the-shelf representation
models that capture token-level evidence and fine-grained semantics.
{\emph{(3)}~We train the proxy with
the oracle's per-document confidence as a soft label, so the proxy
itself learns to be unsure on documents the oracle was unsure
about, instead of being forced toward yes/no only and being confident on documents it
should not be confident on. 
}
\emph{(4)}~Our calibration separates when the proxy is uncertain from when the sample is too small to be sure, and adds a
safety margin only on the latter. This leads to fewer LLM cascades at the same accuracy target.}
{Also, we are the first to use the oracle LLM's own
per-document confidence for three new purposes in this setting:
as a query-level difficulty compass that explains which family
wins on which query, to derive a lower bound on the minimum oracle calls
any proxy-based cascade can make at the target accuracy, and as
the training signal (soft label) for our proxy.}
{At a $90\%$ accuracy target on three
$10\mathrm{K}$-document corpora, our methods are
$1.6$--$2.0\times$ faster than the best prior method on each
corpus, and meet the target accuracy on $95\%$ of queries.
The latency lower bound derived from the oracle confidence
indicates a further $\sim$$4$--$20\times$ of headroom for future
work.}
\end{abstract}


\begin{document}
\maketitle

\label{sec:intro}

\section{Introduction}\label{sec:intro}

Database engines are increasingly asked to evaluate predicates on unstructured data that
are not expressible in SQL on structured data \cite{liu2024palimpzest, patel2024lotus}. A clinician may want every medical publication abstract
that ``describes a randomized trial of an oncology drug''; an analyst
may want every customer review that ``expresses dissatisfaction with
shipping.'' Such predicates have natural-language form, no closed
schema, and no cheap implementation: today the only general way to
evaluate them is to call a Large Language Model (LLM, e.g., \oracleLLM) on every
document~\cite{wang2023zero,scaledoc}. 
This is used as the oracle to avoid more costly and manual human labeling.
Still, on a corpus of 10K
documents a single such predicate already takes thousands of seconds to evaluate exhaustively \cite{scaledoc}; the cost is linear in
the corpus size and document length, i.e., the amount of tokens.

\paragraph{Semantic filtering as a query plan.}
Recent work treats this as a classical database problem: the LLM is
an expensive user-defined function, and the optimizer's job is to
apply it to as few documents as possible while still meeting a
user-specified accuracy target $\alpha$ (e.g.\ $0.90$) \cite{patel2024lotus, liu2024palimpzest, scaledoc}. The dominant
plan is a two-tier \emph{cascade}~\cite{wang2011cascade,bolukbasi2017adaptive,chen2023frugalgpt,geifman}: a lightweight \emph{surrogate or proxy} is
evaluated on every document and the LLM is called only on the
documents the proxy is uncertain about. The plan is parameterized
{by a score threshold that controls how many documents go to the
LLM, and the optimization problem is to pick that threshold (and
the proxy) so the SLA holds at the fewest oracle calls.}

\paragraph{Three families of cascades, none dominant.}
Three lines of work have emerged for the proxy.
\textbf{(i)~Model-free.} \methCSV~\cite{csv2026} clusters the documents, samples a few per cluster, labels them with the oracle
LLM, and propagates the majority label. There is no trained proxy.
\textbf{(ii)~Prebuilt LLM proxy.} \methBARGAIN~\cite{bargain2025} scores
every document with a pre-trained \emph{small} LLM (e.g.\
\proxyLLM) and falls back (or cascades) to the large oracle on the docs the
{small one is least confident about; the score threshold is calibrated on a small labeled set to meet the target accuracy. There is no per-query training.}
\textbf{(iii)~Online-trained proxy.} \methScaleDoc~\cite{scaledoc} trains
a per-query bi-encoder proxy on a small labeled training set with contrastive
learning online.

\begin{figure}[t]
\centering
\includegraphics[width=\linewidth]{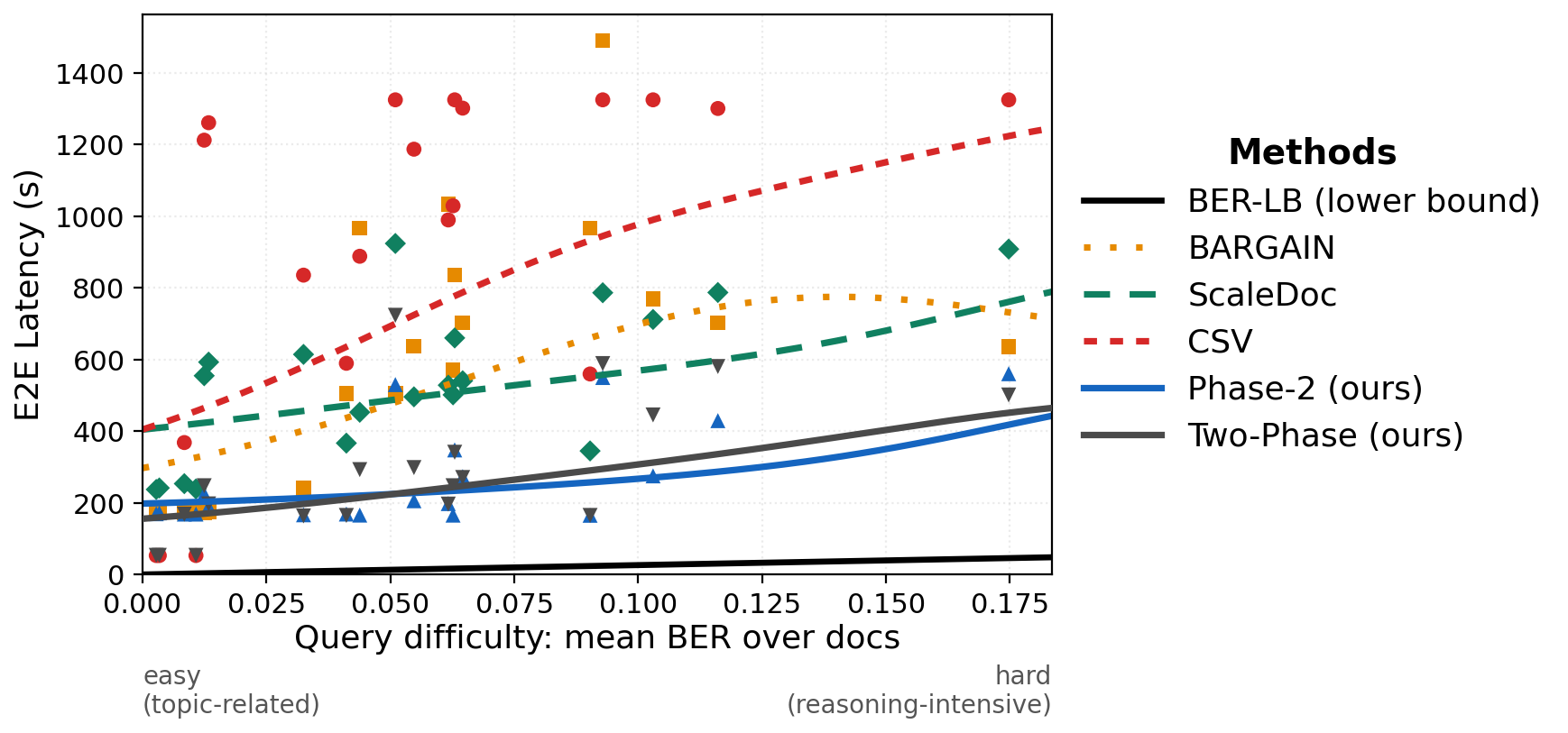}
\caption{
End-to-end latency vs. per-query difficulty (mean Bayes Error Rate (BER) of the oracle LLM over documents), on \dsPubMed corpus at the $90\%$ accuracy target (\S\ref{sec:experiments}).
Each point is a query; each line is a non-parametric smoothing curve. BER-LB is our derivation of optimal performance. All methods cascade to more oracle LLM calls for harder queries to meet accuracy target.}
\label{fig:conceptual_curve}
\end{figure}

Each family has a niche --- \methCSV on topical queries where
embedding clusters align with the predicate; \methBARGAIN on
reasoning-heavy queries where a small LLM already separates yes from
no, but classification with this small LLM adds an overhead; 
\methScaleDoc on the middle ground, with an overhead of online labeling of training set and proxy training --- but no family wins across the query mix we measure. 
Fig.~\ref{fig:conceptual_curve} shows this.
Also, no prior paper explains \emph{which
family wins when}, nor places the families on a single
algorithmic skeleton, and nor expresses \emph{query difficulty as a quantifiable metric}.

\paragraph{A compass: the oracle's own Bayes Error Rate (BER)}
Two facts make a difficulty-aware analysis possible. First, the
oracle LLM returns not only a hard label $y_i$ but also a class
probability $\prob^\star_i$ derived from output token logprobs --- a soft
label that is free in the sense that it requires no extra oracle
call. Second, the per-document Bayes Error
$\eta_i=\min(\prob^\star_i,1-\prob^\star_i)$ is a lower bound on the
expected error of \emph{any} classifier on that document~\cite{fukunaga1990introduction},
because the oracle is the most accurate label we have. The corpus
average $\overline{\BER}_q$ for a query $q$ is therefore a query-level difficulty
score that is method-independent. We use it as a compass to explain which family wins, and to derive a lower
bound on the minimum cascade calls any proxy can make.

\paragraph{Four observations behind the contributions.}
This paper builds on four observations the prior literature has not
isolated. The first three identify limitations in how existing
cascades are designed; the fourth names a calibration that is often either too conservative 
%{optimistic} 
in meeting the target accuracy.

\textbf{(O1) The three families share a skeleton.}
\methCSV, \methBARGAIN, and \methScaleDoc all instantiate the same 
{six-step} template
{(partition, sample, label, build a proxy, calibrate, deploy).} 
%What
%differs is the choice along four orthogonal design knobs:
%proxy representation (embeddings / small-LLM / bi-encoder), training
%(none / prebuilt / online), calibration (none / point-estimate / confidence-bound), 
%and partitioning (none / adaptive
%clustering). 
What
differs is the choice along four orthogonal design knobs:
proxy representation, proxy training, calibration, 
and partitioning. 
The conceptual benefit is large: prior
``methods'' become \emph{cells} in a knobs $\times$ choices
matrix. New methods live in the empty cells or add new choices, and also can be designed through the seamless integration of multiple methods; we propose examples in this paper.

\textbf{(O2) \methScaleDoc{}'s bottleneck is the architecture, not the
training.} 
{\methScaleDoc invests in a custom-designed multi-stage
contrastive training scheme on a simple proxy (bi-encoder over
dense embeddings).} Before designing yet another training scheme,
we revisit traditional representation-learning ML and find the
trade-off is the wrong way around: the bi-encoder's output cosine
similarity is non-separable on \emph{reasoning} predicates, because (a) the dense embedding has already
discarded the token-level evidence the predicate depends on, and
(b) cosine similarity captures only topical relatedness of query
and documents, not more complicated semantics.
{We replace the
architecture with a hybrid of off-the-shelf token-aware
representation models so the proxy \emph{can} distinguish documents
the bi-encoder fuses together; train it with the oracle's
per-document confidence as a soft label, so the proxy itself stays
unsure on documents the oracle was unsure about instead of being
forced toward yes/no;
%and its confidence becomes a faithful estimate of the oracle's; 
and impose a simple training-time constraint that ties training to the user's accuracy target, shipping some of the calibration complexity to training.
}

{\textbf{(O3) Existing calibrations are too conservative,
cascading to the oracle LLM more often than necessary.}
The simplest calibration just uses the proxy's measured error rate
on the labeled sample to set the threshold; we observe this is
\emph{optimistic} on documents the sample does not cover
densely --- a handful of labels can leave the measured error rate
arbitrarily off the truth. %and the threshold 
%leads to calling fewer documents than the accuracy target really allows.
Existing
state-of-the-art calibrations~\cite{scaledoc,bargain2025} fix the
optimism by adding a worst-case safety margin to the measured
error rate. This is safe but \emph{conservative}: the same safety margin
is added on documents the labeled sample has covered densely
\emph{and} on documents it has not. 
%calling the LLM on documents
%the labeled sample has already pinned down. 
We propose a
calibration that adds a safety margin only on documents where the
labeled sample is small, so the threshold tightens on the
densely-covered documents and stays safe on the rest.}

\textbf{(O4) No single method wins on all queries.}
Each family is the fastest plan in a different region of query
difficulty: \methCSV on easy queries, our proxy on medium- to hard
queries. 
The cheapest deployable plan would route each query to its best family. 
However, the routing decision itself costs oracle labels — a naive router spends a sample to identify the winner and then throws those labels away.
We propose an adaptive \emph{\methTwoPhase} method that tackles this: it runs \methCSV as Phase~1, terminates
early when Phase~1 is efficient enough to classify all documents confidently, and otherwise reuses Phase-1's
oracle labels as Phase-2 training data for our proxy. 
%The transition between phases adds no extra oracle calls.

\paragraph{Contributions.}
{We make the following contributions (C1--C5).}
%organized so they can be adopted
%independently.

\begin{enumerate}[topsep=2pt,itemsep=2pt,leftmargin=1.6em]
\item \textbf{A unified framework}
  (\S\ref{sec:semfilt}--\S\ref{sec:semfilt:unified}) places \methCSV,
  \methBARGAIN, \methScaleDoc, and our methods on one 
  {six-step skeleton}
  with four explicit design knobs. The framework converts ``different
  methods'' into ``different cells of one design matrix''.
\item \textbf{A better online proxy by revisiting traditional ML}
  (\S\ref{sec:proxy}). 
  {We
  identify the bottleneck in \methScaleDoc as the proxy's
  \emph{architecture}, not its training. We replace it with a
  hybrid of off-the-shelf token-aware representation models that
  capture richer query--document semantics than cosine over dense
  embeddings; train it with the oracle's per-document confidence
  as the soft label; and impose a small training-time constraint that aligns the
  proxy's confidence with the deployment-time accuracy target.}
{\item \textbf{A tighter calibration}
  (\S\ref{sec:bercp}). 
  %A single-knob refinement of the calibration used in prior cascades. 
  We increase the safety margin only on
  documents the labeled sample is too small to cover, and leave the safety margin at zero on documents the labeled sample
  already covers; our threshold calls the oracle LLM less often than
  prior calibrations at the same accuracy target. %without ever trusting a labelled set that is too small to be trusted.
  }
\item \textbf{Adaptive two-phase method} (\S\ref{sec:twophase}). A method
  that combines \methCSV (Phase~1) with our proxy and calibration
  (Phase~2), reusing Phase-1 oracle labels as Phase-2 training data to minimize the transition cost. Phase~2 is bypassed when
  Phase~1 is enough to classify all documents on easy queries.

{\item \textbf{BER as compass and headroom}
(\S\ref{sec:headroom}). Cross-cutting across (C1)--(C4): we use
the oracle's own BER both as a query-level difficulty score
(explaining which family wins on which query) and to derive a
lower bound on the minimum LLM cascade calls any proxy-based
method can make at the target accuracy.}
{Our soft-label training (C2) reuses the same per-document
confidence as BER.}

\end{enumerate}

\paragraph{Headline empirical result.}
{On three datasets of $10$K documents and $20$ queries each at a
$90\%$ accuracy target, our two-phase method runs $1.6$--$2.0\times$
faster end-to-end than the best prior method per corpus, is at or
near the fastest plan on every query, and meets the target
accuracy on $95\%$ of queries. The
BER-derived latency lower bound sits a further $\sim$$4$--$20\times$
below it, indicating that semantic-filter optimization still has
substantial headroom for future work.}

\paragraph{Plan of the paper.}
\S\ref{sec:related} surveys related work.
\S\ref{sec:semfilt} defines the problem and introduces the unified framework (C1). \S\ref{sec:proxy}--\ref{sec:twophase}
present the three technical contributions:
the revisited proxy (C2), calibration (C3), and the
two-phase method (C4). \S\ref{sec:headroom} develops BER as a
difficulty compass and a lower bound on cost (C5).
\S\ref{sec:experiments} reports empirical results, and
\S\ref{sec:conclusion} concludes.

%\label{sec:related}
%\ref{tab:positioning}
%\cite{lu2023probabilistic}

\section{Related Work}\label{sec:related}

This section explains the closest research areas with representative and state-of-the-art studies.
%We place this paper between five lines of work and end with a
%property matrix (Table~\ref{tab:positioning}) that compares
%prior systems against the goals a semantic-filter user actually
%cares about.

\paragraph{LLM-powered data systems.}
Treating an LLM call as a user-defined function inside a relational
plan goes back to probabilistic predicates over rich
media~\cite{lu2023probabilistic} and continues with the recent wave
of {LLM-powered data systems} such as
\methPalimpzest~\cite{liu2024palimpzest}, \methLOTUS~\cite{patel2024lotus}, and
\methDocETL~\cite{shankar2024docetl}. {These supply a query
language (e.g., semantic \textsc{Where}/\textsc{Group By}/\textsc{Join}) and
an optimizer that performs both \emph{logical} optimization (operator
reordering, plan rewrites) and \emph{physical} optimization (selecting the best one among existing physical operators, or proxies in the fast-proxy-then-oracle-LLM cascade to reduce expensive oracle-LLM calls). Concurrent papers each refine a particular layer of this
stack: \methPLOP~\cite{mang2026plop} extends the \emph{logical} layer with
a dynamic-programming placement of semantic operators relative to
relational ones; Chung et al.~\cite{chung2026proxy} characterize
lightweight proxies as a class of \emph{physical} operators for
AI-extended SQL warehouses; and KEN~\cite{kossmann2026ken} sits
\emph{below} the optimizer at the execution layer, scheduling
cascade models on the GPU.}

{Our work occupies a different point in this stack, proposing a \emph{new physical implementation} of the semantic filter
operator that the LLM-powered data systems above can employ.
\methSUPG~\cite{kang2020supg} used the cascading idea in the pre-LLM era, alike the speculation in database query processing~\cite{sioulas2021speculation}. \methLOTUS{} and \methPalimpzest{} above adopted it
as one of their LLM-based physical implementations, and the focused state-of-the-art methods---our direct baselines---are
\methScaleDoc~\cite{scaledoc}, \methCSV~\cite{csv2026}, and
\methBARGAIN~\cite{bargain2025}, discussed next. 
To our knowledge, no prior work has compared these state-of-the-art
operators head-to-head in a single experimental
setting.}

\paragraph{New physical operator for semantic filter.}
The closest line is cascades that pair a cheap proxy score with
the oracle. They are the prior art our contributions sit alongside.
%so we describe each carefully.

{\textbf{\methScaleDoc}~\cite{scaledoc} embeds documents
into vectors offline, trains a per-query bi-encoder proxy
(taking document and query embeddings as inputs and outputting a
cosine similarity of transformed embeddings) \emph{online} via a
multi-stage contrastive scheme on a small oracle-labeled training
sample. At deployment, it draws a stratified calibration sample,
builds a $64$-bin histogram of yes/no counts on the cosine score,
smooths the per-bin counts, and searches for a two-sided band
$[\ell,u]$ on the cosine score whose expected accuracy under the
smoothed per-bin estimates meets the user target. Documents whose
score falls outside the band are auto-labeled; documents inside
the band are cascaded to the oracle. While its proxy adapts to
each query online, (i)~the bi-encoder over embeddings captures
only topical similarity, leaving token-level evidence out of
reach; and (ii)~the calibration smooths the per-bin counts on the labeled sample before setting the threshold, which is a deliberate safety choice that prevents an optimistic threshold on score ranges the labeled sample covers only sparsely; but the same smoothing widens the auto-cascade band on \emph{every} score range, including the well-covered ones where the measured error rate is already reliable on its own, so the threshold ends up calling the LLM on more documents than necessary to meet the target accuracy.}

\begin{table*}[t]
\centering
\small
\caption{\textbf{Desired properties in semantic filter.} \ding{51}:
delivered; \ding{55}: not delivered; $\sim$: partially.
Explanation of $\sim$ marks:
(\methCSV{}) not declarative in the original paper, but
setting internal parameters trivially meets target accuracy;
(\methBARGAIN{}) prebuilt small LLM is
partially trustworthy to oracle LLM; it processes tokens, but its scan over all documents adds a substantial cost.}
\label{tab:positioning}
\begin{tabularx}{\textwidth}{Xcccc}
\toprule
\textbf{User-facing goal} & \textbf{\methScaleDoc} & \textbf{\methCSV} & \textbf{\methBARGAIN} & \textbf{Ours} \\
\midrule
\emph{Declarative target accuracy.} No need to manually tune internal parameters to meet target accuracy. & \ding{51} & $\sim$ (no guide) & \ding{51} & \ding{51} \\
\addlinespace
\emph{Headroom diagnosis.} Measure how close the latency is to the
optimal. & \ding{55} & \ding{55} & \ding{55} & \ding{51} \\
\addlinespace
\emph{Trustworthy confidence.} The proxy's score reflects how likely
the answer is to be correct. & \ding{55} & \ding{55} & $\sim$ (small LLM) & \ding{51} \\
\addlinespace
\emph{Expressive predicates.} Efficiently support queries with token-level
evidence beyond topical similarity. & \ding{55} & \ding{55} & $\sim$ (small LLM) & \ding{51} \\
\addlinespace
\emph{Tight calibration.} The proxy score threshold is not too optimistic or too
conservative. & \ding{55} & \ding{55} & \ding{55} & \ding{51} \\
\addlinespace
\emph{Per-query competitiveness.} A single method is
{competitive} with the per-query best on every query.
& \ding{55} & \ding{55} & \ding{55} & \ding{51} \\
\bottomrule
\end{tabularx}
\end{table*}

{\textbf{\methCSV}~\cite{csv2026} is a \emph{model-free}
cascade: it clusters documents in dense embedding space, asks the
oracle to label a small per-cluster sample, and propagates the
majority label to the entire cluster if the per-cluster sample
agrees on at least a vote-threshold fraction of its documents
(\methCSV{} leaves this threshold as a free hyper-parameter; in
our experiments we set it equal to the user's target accuracy
$\alpha$ -- see \S\ref{sec:twophase:design}). Clusters whose sample does not agree are}
re-clustered into smaller ones; persistent disagreement falls
back to per-document oracle calls. \methCSV is cheap when the
embedding clustering aligns with the predicate (so most clusters
are nearly all positive or all negative) but expensive
when it does not: the same token-level evidence the
\methScaleDoc bi-encoder discards is also missing from the
embeddings \methCSV clusters on, so predicates that depend on
that evidence end up with many mixed clusters and a large oracle call cost. 

%in-cluster label
%disagreement is high.

\textbf{\methBARGAIN}~\cite{bargain2025} differs from \methScaleDoc in two ways. The proxy is a prebuilt small LLM rather than an online-trained bi-encoder, so the calibration sample is the only labeling cost beyond the cascade itself, without a separate training sample. The calibration replaces \methScaleDoc{}'s smoothed empirical rate with a distribution-free high-confidence upper bound per score interval; this is finite-sample valid but \emph{more} conservative than \methScaleDoc{}'s smoothing on every well-covered interval, because the upper bound inflates the per-interval error rate above what the smoothed estimate would --- so the threshold calls the LLM more often than \methScaleDoc does, and substantially more often than the target accuracy requires. The per-document small-LLM scan over the whole corpus adds a moderate cost.

\paragraph{What a semantic filter should offer its users.}
Table~\ref{tab:positioning} states six desired properties
and marks which methods deliver each. (1) Prior methods deliver
declarative accuracy targets, but none of them lets the user
(2) see how close cost is to the optimal performance bound,
(3) trust the proxy's score as reflecting how likely the answer is
to be correct, (4) answer queries that go beyond topical
relatedness without calling the oracle frequently, (5) get a tight proxy-score threshold that meets target accuracy
without being too optimistic (SLA violated) or too conservative
(oracle over-called), or
(6) deploy a single method that is \emph{competitive} with
the per-query best on \emph{every} query, without having to know or
hand-pick which method is best in advance. To the best of our knowledge, our work is the first to deliver all six.

\section{Semantic Filtering and a Unified Framework}\label{sec:semfilt}

This section states the problem (\S\ref{sec:semfilt:problem}), cost model
(\S\ref{sec:semfilt:cost}), and then unifies prior work on a single
algorithmic framework (\S\ref{sec:semfilt:unified}).

\subsection{Problem}\label{sec:semfilt:problem}

A corpus $\Dpool=\{d_1,\dots,d_N\}$ and a natural-language predicate or query
$q$ are given. A designated LLM (the \emph{oracle}) returns, for any
$(q,d_i)$, a hard label $y_i\in\{0,1\}$ optionally with a class probability $\prob^\star_i = P(y_i = yes) \in[0,1]$ derived from token logprobs. A semantic filtering method outputs predicted labels
$\hat y_1,\dots,\hat y_N$ such that the corpus accuracy
$\acc=\tfrac{1}{N}\sum_i\mathbf{1}[\hat y_i=y_i]$ is at least a
user-specified target $\alpha$ (e.g., 0.9). 
% with failure probability at most $\delta$ over the algorithm's internal randomness.

Throughout the paper we treat the oracle as the ground truth. This is
the standard convention in semantic-filter
work~\cite{scaledoc,csv2026,bargain2025}, and any disagreement
between the oracle and an underlying notion of ``truth'' is folded
into the irreducible noise we measure in \S\ref{sec:headroom}.

\subsection{Cost Model}\label{sec:semfilt:cost}

We adopt the \methScaleDoc-style end-to-end accounting~\cite{scaledoc} and
refine it, assuming the proxy-then-oracle-LLM pipeline. The deployable cost has two terms:
\begin{equation}\label{eq:cost}
  C \;=\; \underbrace{\Tpx(n_{\mathrm{tr}}, N)}_{\text{proxy train/score time}}
  \;+\;\underbrace{\bigl(n_{\mathrm{tr}}+n_{\mathrm{ca}}+n_{\mathrm{cas}}\bigr)\,\tllm}_{\text{oracle calls}}.
\end{equation}
Here $\tllm$ is the average oracle latency per call (measured on the
deployment GPU); $n_{\mathrm{tr}}$ and $n_{\mathrm{ca}}$ are the
training and calibration set sizes; $n_{\mathrm{cas}}$ is the
cascade count after proxy deployment, calling the oracle for the docs the proxy is uncertain about;
and 
$\Tpx(n_{\mathrm{tr}}, N)$ is the time for proxy training and scoring documents with proxy.
%$\Tpx(n_{\mathrm{tr}}) \approx \alpha_0 + \alpha_1 n_{\mathrm{tr}} + S(N - n_{\mathrm{tr}})$ with $\alpha_1$ the per-label training-cost slope and $S(N - n_{\mathrm{tr}})$ the cost of scoring all $N - n_{\mathrm{tr}}$ documents. Empirically (\S\ref{sec:experiments}) $\Tpx$ is linear in $n_{\mathrm{tr}}$ with $R^2\in[0.94,0.98]$ across the queries on \dsPubMed.

\paragraph{The oracle's soft label~\cite{hinton2015distilling,pereyra2017regularizing} is free.}
Every modern LLM API returns the logprobs of its output tokens, so $\prob^\star_i$ is exposed at no
extra cost. Throughout the paper we exploit it as a soft label to train our proxy and compute the BER.

\subsection{A Unified Algorithmic Framework}\label{sec:semfilt:unified}

{The three families of cascades in
\S\ref{sec:related} share the same six-step skeleton, stated
abstractly in Algorithm~\ref{alg:unified} and drawn as a DAG in
Figure~\ref{fig:framework}. Methods differ only in the choices
they make for four orthogonal design knobs, summarized
in Figure~\ref{fig:framework_matrix}.}

\begin{algorithm}[t]
\caption{A unified cascade framework for semantic filtering.}
\label{alg:unified}
\begin{algorithmic}[1]
\Input Corpus $\Dpool$, predicate $q$, accuracy target $\alpha$
\Output Predicted yes/no labels
\State \textbf{Partition.} Group the documents (e.g.\ by embedding
clustering or trivially into one group).
\State \textbf{Sample.} Draw a small labeled set per group.
\State \textbf{Label.} Call the oracle on the sample; collect hard
labels $y$ and optionally soft labels $\prob^\star$.
\State \textbf{Build a proxy.} One of (a)~no model, (b)~train an online proxy on the labeled training set, (c)~reuse a prebuilt
small-LLM proxy.
\State \textbf{Calibrate.} 
{Choose a score threshold $\tau$ (or a decision rule) on a
held-out portion of the labeled calibration set so that the SLA with $\alpha$ is expected to hold.}
\State \textbf{Deploy.} %Rewritten: lead with the per-document decision; recursion is the last clause and points to the Fig.~2 arrow.
First check if the partition's deploy condition fails (e.g.\ \methCSV{}'s vote does not agree), recurs to step~1 to re-partition (the dashed back-arrow in Figure~\ref{fig:framework}). Then, for each remaining document, auto-label or cascade it to the oracle based on its proxy score and the calibration.\\
\Return per-document predictions.
\end{algorithmic}
\end{algorithm}

%% The "Four design knobs" header + bullet list moved below the DAG
%% paragraph so the section narrates Alg.~1 + Fig.~3 first, then the
%% knob list with Fig.~2.

{%
\paragraph{The framework as a DAG}
Figure~\ref{fig:framework} draws Algorithm~\ref{alg:unified} as a
six-node DAG. The framework makes one structural
observation natural: the oracle labels collected in step~3 are
\emph{the same kind of object} regardless of which method
collected them, so they can also be used as the training input
of a \emph{different} method's proxy. The dashed green arrow marks this cross-method join: two families can share a
single labeling pass and a simple decision condition (e.g.\
``once $\sim$$7\%$ of the corpus is labeled and the first
method is not confident enough, hand off to the second method'')
can switch between them. Our \methTwoPhase composes \methCSV
and our proxy this way (\S\ref{sec:twophase}).}

{\paragraph{Four design knobs.}
The differences between \methCSV, \methBARGAIN, \methScaleDoc,
and our \methBERCP and \methTwoPhase reduce to four orthogonal
choices, displayed in
Figure~\ref{fig:framework_matrix}. 
We list its high-level options here; the per-method choices and the details behind them are in \S\ref{sec:related} and
\S\ref{sec:proxy}--\ref{sec:twophase}.}

{%
\begin{itemize}[topsep=2pt,itemsep=2pt,leftmargin=2em]
\item \textbf{Representation.} How does the proxy compute a
  document's score? Options include using documents' \emph{dense embeddings} without a special ML model, \emph{an ML model over query and document}, or a
  \emph{small LLM}.
\item \textbf{Training.} Is the score function learned
  per-query, used as-is from a pre-training, or
  not learned at all?
\item \textbf{Calibration.} How is the threshold that meets the
  accuracy target chosen on the labeled sample? Choices range
  from a simple vote-agreement threshold set to $\alpha$ to
  high-confidence upper bounds on the proxy's error rate.
\item \textbf{Partition granularity.} Are documents grouped by \emph{embedding clustering} or left as a
  \emph{single group}?
\end{itemize}}

{%
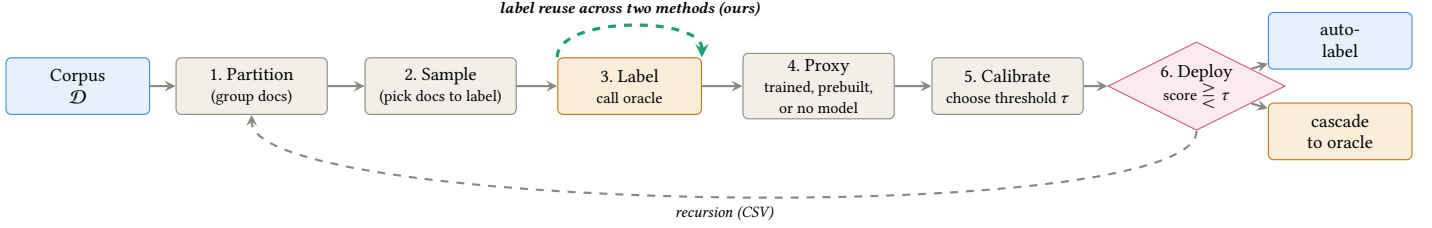
\begin{figure*}[t]
\centering
\definecolor{cdoc}{HTML}{E6F1FB}
\definecolor{cdocedge}{HTML}{378ADD}
\definecolor{coracle}{HTML}{FAEEDA}
\definecolor{coracleedge}{HTML}{BA7517}
\definecolor{cdec}{HTML}{FBEAF0}
\definecolor{cdecedge}{HTML}{D4537E}
\definecolor{cop}{HTML}{F1EFE8}
\definecolor{copedge}{HTML}{888780}
\definecolor{cours}{HTML}{E1F5EE}
\definecolor{coursedge}{HTML}{1D9E75}
\begin{tikzpicture}[
  font=\footnotesize,
  node distance=8mm,
  every node/.style={align=center},
  docset/.style={draw=cdocedge, fill=cdoc, rounded corners=2pt, minimum width=1.9cm, minimum height=0.75cm, inner sep=2pt},
  oracle/.style={draw=coracleedge, fill=coracle, rounded corners=2pt, minimum width=1.9cm, minimum height=0.75cm, inner sep=2pt},
  op/.style={draw=copedge, fill=cop, rounded corners=2pt, minimum width=2.0cm, minimum height=0.75cm, inner sep=2pt},
  dec/.style={draw=cdecedge, fill=cdec, diamond, aspect=2, inner sep=1pt, minimum width=1.7cm},
  arr/.style={->, >=stealth, thick, draw=copedge},
  share/.style={->, >=stealth, very thick, dashed, draw=coursedge},
  recur/.style={->, >=stealth, thick, dashed, draw=copedge},
  slabel/.style={font=\scriptsize\bfseries, anchor=south}
]
% Row of six framework steps left to right
\node[docset] (corpus) at (0, 0) {Corpus\\$\Dpool$};
\node[op]     (part)   at (2.3, 0) {1.~Partition\\\scriptsize (group docs)};
\node[op]     (samp)   at (4.8, 0) {2.~Sample\\\scriptsize (pick docs to label)};
\node[oracle] (label)  at (7.3, 0) {3.~Label\\\scriptsize call oracle};
\node[op]     (build)  at (9.8, 0) {4.~Proxy\\\scriptsize trained, prebuilt,\\\scriptsize or no model};
\node[op]     (cal)    at (12.3, 0) {5.~Calibrate\\\scriptsize choose threshold $\tau$};
\node[dec]    (dep)    at (14.8, 0) {6.~Deploy\\\scriptsize score $\gtreqless \tau$};

\draw[arr] (corpus) -- (part);
\draw[arr] (part)   -- (samp);
\draw[arr] (samp)   -- (label);
\draw[arr] (label)  -- (build);
\draw[arr] (build)  -- (cal);
\draw[arr] (cal)    -- (dep);

% Output legs from Deploy
\node[docset] (auto) at (16.7, 0.6) {auto-\\label};
\node[oracle] (casc) at (16.7,-0.6) {cascade\\to oracle};
\draw[arr] (dep) -- (auto);
\draw[arr] (dep) -- (casc);

% Optional recursion: the arrow exits from the same side of Deploy as
% the Deploy -> cascade arrow (the score <= tau branch), curves below
% the row, and returns to Partition. Triggered when the auto-label
% condition fails on a partition (e.g. CSV's vote does not agree
% strongly enough and the cluster is re-clustered before any
% individual document is cascaded to the oracle).
\draw[recur] (dep.south) .. controls +(0, -1.3) and +(0, -1.3) .. (part.south)
  node[midway, below, font=\scriptsize\itshape]
       {recursion (\methCSV)};

% Cross-method join (our contribution): dashed self-loop ABOVE the
% Label node (was above Sample). Tightened controls so the head and
% root are close together.
\draw[share] (label.north west) .. controls +(0, 0.55) and +(0, 0.55) .. (label.north east)
  node[midway, above, font=\scriptsize\itshape]
       {\textbf{label reuse across two methods (ours)}};

\end{tikzpicture}
\caption{
{\textbf{The unified framework as a DAG.}
The six steps of Algorithm~\ref{alg:unified} arranged left to
right. Blue boxes are document sets; amber boxes are
oracle-labeled subsets; gray boxes
are operations; the pink diamond is the deploy-time decision
(auto-label vs.\ cascade) gated by whether the proxy score
passes the calibration threshold $\tau$. 
The dashed black arrow marks a recursion when the deploy condition
fails on a partition (e.g.\ \methCSV{} re-clusters a cluster
whose vote does not agree strongly enough, cascading individual
documents to the oracle only as a last resort).
The dashed green arrow is \textbf{our
cross-method join}: the same labeled sample drawn for one
method can be reused as the labeled sample of a second method (\S\ref{sec:twophase}).}}
\label{fig:framework}
\end{figure*}}

% Matrix rendered as a vector PDF by figures/make_framework_matrix.py
% (matplotlib). Switching off of LaTeX \cellcolor avoids the white-sliver
% / uneven-row-height artefacts that \arraystretch, \extrarowheight, and
% \rule struts could not fully suppress on this template.
\begin{figure*}[t]
\centering
\includegraphics[width=\textwidth]{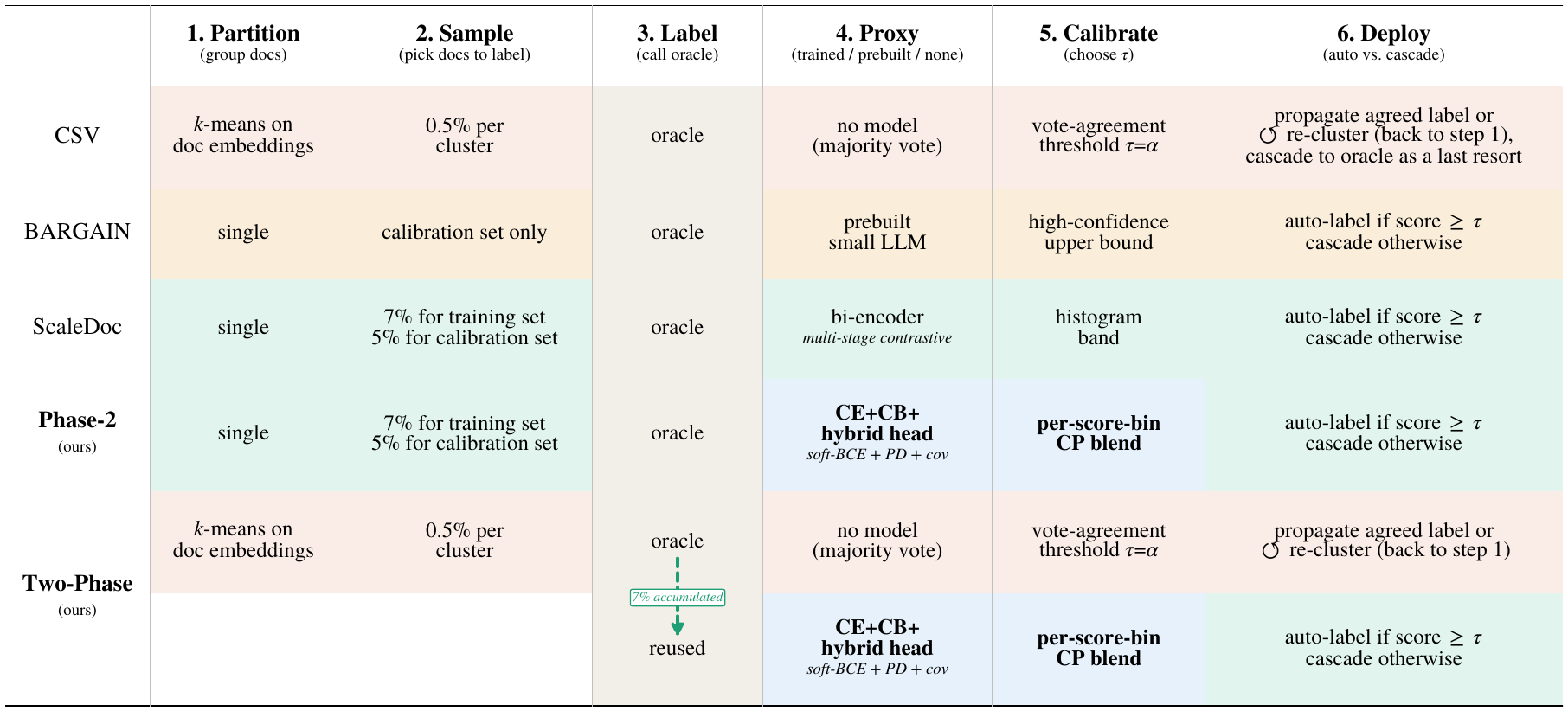}
\caption{\textbf{Method-by-step matrix view.} Rows are methods; columns are the six steps. Cell color identifies the family (red = \methCSV, amber = \methBARGAIN, teal = \methScaleDoc); \textbf{blue cells are our contributions}. \methCSV{}'s Deploy cell contains the recursion ($\circlearrowleft$) that matches the dashed back-arrow in Figure~\ref{fig:framework}. \methTwoPhase is shown as two sub-rows: the upper (\methCSV-style) first phase, the lower (\methBERCP-style) second phase; the green arrow is the label reuse in Figure~\ref{fig:framework} and fires when $\sim$$7\%$ of the corpus has been labeled.}
\label{fig:framework_matrix}
\end{figure*}

{%
\paragraph{Our contributions on the matrix.}
The blue-cell rows in Figure~\ref{fig:framework_matrix} mark
the cells unoccupied by prior work. \methBERCP introduces
a new proxy in step~4 (cross-encoder (CE) + \ColBERT (CB) + hybrid head over them,
\S\ref{sec:proxy}) and a new calibration in step~5 (per-score-bin
upper-bound blend, \S\ref{sec:bercp}); it keeps the trivial
single-group partition and the random sample of \methScaleDoc.
\methTwoPhase (row~5) uses \methCSV as the first phase -- clustering and per-cluster sample in steps~1--2, with the labels accumulated in step~3 over re-clustering (step 6 to 1). When $\sim$$7\%$ of corpus is labeled in step~3, it reuses the labeled docs to train our proxy (\S\ref{sec:twophase}).
}
\section{A Better Online Proxy by Revisiting Traditional ML}\label{sec:proxy}

This section shows that the bottleneck in online-proxy approach is 
the proxy's \emph{architecture}, more than its training
(\S\ref{sec:proxy:diag}), and presents our replacement: {a
token-aware hybrid (\S\ref{sec:proxy:arch}) trained with binary
cross-entropy on the oracle's soft labels and a small SLA-aware
primal--dual constraint (\S\ref{sec:proxy:train}).}

\subsection{Diagnosis: \methScaleDoc's suboptimal proxy}\label{sec:proxy:diag}

\methScaleDoc{}'s proxy is a \emph{bi-encoder}~\cite{reimers2019sentence,karpukhin2020dense,devlin2019bert}. Query and document each pass
through a frozen embedding model, then a $55$M-parameter MLP
projects each to another dense representation, and the score is the cosine similarity.

\begin{figure}[t]
\centering
\begin{subfigure}[b]{0.31\linewidth}
  \centering
  \includegraphics[width=\linewidth]{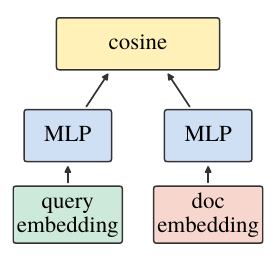}
  \caption{Bi-encoder}
  \label{fig:proxy-bi}
\end{subfigure}
\hfill
\begin{subfigure}[b]{0.31\linewidth}
  \centering
  \includegraphics[width=\linewidth]{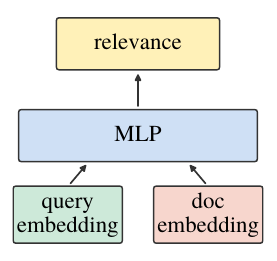}
  \caption{Cross-encoder}
  \label{fig:proxy-ce}
\end{subfigure}
\hfill
\begin{subfigure}[b]{0.31\linewidth}
  \centering
  \includegraphics[width=\linewidth]{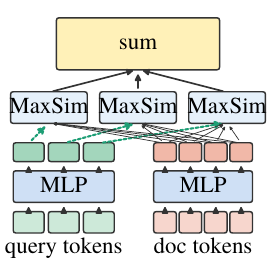}
  \caption{ColBERT}
  \label{fig:proxy-cb}
\end{subfigure}
% Removed leftover outer {...} that were left behind when this \caption{} was
% group that locally swallowed \@currentlabel, leaving the figure's \label
% pointing at the last subfigure (Reference fig:proxy-architectures
% undefined on page 12 in the previous compile log).
\caption{\textbf{Three model architectures.}
\textbf{(a)~Bi-encoder}: query and document embeddings are
projected by two independent MLPs; the cosine of the projected
vectors is the relevance score.
\textbf{(b)~Cross-encoder:} the two embeddings are read jointly by
one MLP that attends across them, producing a single relevance
score.
\textbf{(c)~ColBERT:} query and document \emph{tokens} are encoded
independently by two MLPs into per-token output cells; for each
query output cell, MaxSim picks the largest similarity against any
document output cell, and the per-token MaxSim values are summed
into the final relevance score. Our proxy combines (b) and (c)
through a small hybrid head, which turns the two relevance scores
into the predicted probability $\prob_i$ (and thus the cascade
score $\score_i = 2|\prob_i - \tfrac{1}{2}|$).}
\label{fig:proxy-architectures}
\end{figure}

{\methScaleDoc invests in a multi-stage contrastive training~\cite{chen2020simclr,gao2021simcse}
mechanism on this bi-encoder.} However, dense embeddings discard token-level evidence, and cosine similarity over dense representations approximates \emph{topical} similarity only, %and topical similarity is not a monotone function of the predicate's
%conditional probability for 
which makes it challenging to support reasoning-intensive predicates.
%predicate that depends on reasoning,
%negation, or roles. 
Two documents that are nearly identical
topically can have opposite labels 
two documents far apart in cosine space can both be yes.
Contrastive training cannot fix what the representation has discarded.

\begin{figure}[t]
\centering
\includegraphics[width=0.95\linewidth]{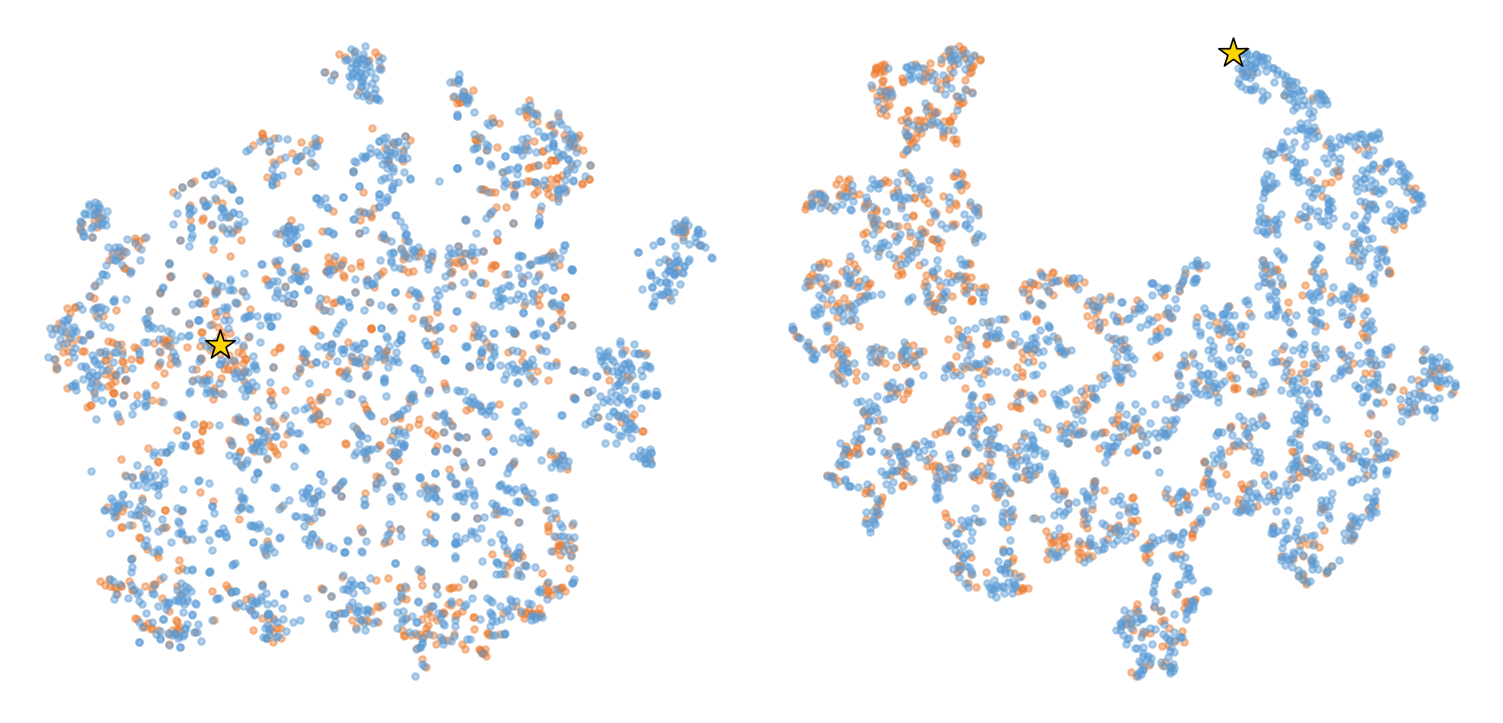}
\caption{T-SNE visualization on raw embedding space (left) and bi-encoded space (right), with blue dots = yes-docs, orange dots = no-docs, star = query, for a \dsPubMed query over $\sim$3K sampled docs.}
\label{fig:tsne}
\end{figure}

\subsection{Architecture: CE $+$ CB $+$ hybrid head}\label{sec:proxy:arch}

{Before describing the components, we name the ideas we are
borrowing and why. Two scoring architectures from neural
information retrieval (IR) --- the \emph{cross-encoder} (CE), popularized
for passage re-ranking by Nogueira and Cho~\cite{nogueira2019passage,lin2021pretrained},
and \emph{ColBERT} (CB), introduced by Khattab and Zaharia~\cite{khattab2020colbert,santhanam2022colbertv2} (Figure \ref{fig:proxy-architectures})
--- have been the textbook examples of outperforming a bi-encoder on
accuracy at IR benchmarks (e.g., MS MARCO~\cite{bajaj2016msmarco}, TREC passage ranking~\cite{craswell2020trec,thakur2021beir}).
A CE captures cross-query-doc interactions via a shared MLP, 
while CB keeps a separate vector per token on each side and aggregates similarity at the token level
(called \emph{late interaction}). The trade-off is well-known: a
bi-encoder reduces each side to one fixed-size vector that is cheap
to compare, but the compression is lossy on any predicate whose
answer hinges on a specific token (a negation, a named entity, a
number); CE captures more complex interactions, and CB recovers that token-level
evidence at the cost of higher per-document scoring.

Neither CE nor CB have, to our knowledge, been used
as the proxy in the LLM-based semantic filtering. 
Rather than redesigning yet another proxy, we \emph{revisit} these mature IR ideas in this setting: our proxy runs both CE and CB, and fuses the two scores with a small
head.}
Also in this setting, the bottleneck is mostly on the model training, not scoring, and we can keep this training overhead small by limiting the model size, in total $\sim$$6\times$ smaller than \methScaleDoc{}'s encoder.

%We replace the bi-encoder with a smaller three-component architecture that reads
%query and document jointly. Total proxy size is $\sim$$9.6$\,M
%parameters, $6\times$ smaller than \methScaleDoc{}'s encoder.

\begin{enumerate}[topsep=2pt,itemsep=3pt,leftmargin=2em]
\item \textbf{\CrossEncoder (CE)} ($\sim$$9.5$\,M).
  An MLP takes query and document embeddings together and outputs a single score $\score^{\mathrm{ce}}$. CE captures cross-interaction between query and document unlike the separate MLPs in the bi-encoder.
\item \textbf{\ColBERT-style projection (CB)} ($\sim$$0.10$\,M).
  Late-interaction scoring: project document and query \emph{tokens} into a
  shared space and aggregate per-document by max-similarity
  over tokens (\textsc{MaxSim}). The CB score $\score^{\mathrm{cb}}$
  is complementary to CE because it weights individual matching
  tokens (negation cues, named entities, numbers) rather than the
  overall fusion.
\item \textbf{Hybrid head} ($\sim$$1.3$\,K).
  A small MLP on a six-dimensional feature vector
  $X_i = [\score^{\mathrm{ce}}_i,\,\score^{\mathrm{cb}}_i,\,
  \score^{\mathrm{ce}}_i \score^{\mathrm{cb}}_i,\,
  |\score^{\mathrm{ce}}_i - \score^{\mathrm{cb}}_i|,\,
  (\score^{\mathrm{ce}}_i)^2,\,(\score^{\mathrm{cb}}_i)^2]$
  produces the proxy's
  {predicted probability
  $\prob_i = \sigma(\mathrm{MLP}(X_i))$ for document $d_i$. The
  cascade in \S\ref{sec:bercp}--\S\ref{sec:twophase} thresholds
  the derived \emph{score}
  $\score_i = 2\,|\prob_i - \tfrac{1}{2}| \in [0,1]$ -- large when
  the proxy is certain (either yes or no), small when the proxy is
  uncertain.}
  %The engineered features make
  %the head trivially small and stable to train on a few hundred
  %labels.
\end{enumerate}

%The composite proxy thus reads query and document jointly twice (CE and CB) and fuses the scores by a learned head. 
%Its capacity is much smaller than a bi-encoder's hidden projection because the bi-encoder
%must encode all of language; the CE and \ColBERT projections are
%pre-trained relevance scorers that only need a thin per-query head.

\subsection{Training: soft-label BCE with SLA-aware primal--dual constraint}\label{sec:proxy:train}

{Our training loss has three terms to answer three questions.
%Each term answers a question \methScaleDoc{}'s contrastive scheme answers indirectly or not at all. 
%We state the question first, then the term that answers it. 
Throughout, $\Tset$ is the labeled training sample
($\sim$$7\%$ of corpus), and the CE, CB, and hybrid head are trained separately for 60, 15, and 120 epochs each.}

{\paragraph{(a) Soft-label cross-entropy: how should the
proxy's probability relate to the oracle's probability?}
{The auto-label vs.\ cascade calibration in \S\ref{sec:bercp} relies on
the predicted probability $\prob_i$ to decide how likely document
$d_i$ is to satisfy the predicate. Training with hard labels
$y_i\in\{0,1\}$ forces the proxy toward $\prob_i\in\{0,1\}$ on
\emph{every} training document, including the ones where the
oracle itself was unsure: the proxy might become confident on
documents the oracle was not, and the calibration cannot tell
``proxy is right'' apart from ``proxy is overconfident.''
We instead use the oracle's per-document probability
$\prob^\star_i\in[0,1]$ as the training target:
\begin{equation}\label{eq:soft-bce}
  \mathcal{L}_{\text{soft}}
  \;=\;\frac{1}{|\Tset|}\sum_{d_i\in\Tset}
  \mathrm{BCE}\bigl(\prob_i,\,\prob^\star_i\bigr),
\end{equation}
where $\mathrm{BCE}(\hat p, p)=-p\log\hat p - (1-p)\log(1-\hat p)$
is binary cross-entropy with a continuous target. At convergence,
$\prob_i\approx\prob^\star_i$ on every training document: the
proxy's predicted probability tracks the oracle's. On documents
the oracle was unsure about, $\prob_i$ stays near $\tfrac{1}{2}$,
so the derived score $\score_i = 2|\prob_i-\tfrac{1}{2}|$ stays
small and the calibration cascades them to the oracle.}
The ablation in \S\ref{sec:experiments:rq6} confirms that this
matters at the deployment cost: hard-label BCE on the same
architecture regresses by 
{$\sim$$8\%$} on end-to-end cost.
}

{\paragraph{(b) SLA-aware constraint: how do we keep the
proxy honest at the deployment target?}
{The proxy's training loss does not know the deployment target accuracy
$\alpha$. Without a coupling between the two, training can leave
the proxy at a calibration that the deploy-time threshold then has
to fix by sending more documents to the LLM. We mitigate this by
adding a single soft constraint to the training loss, evaluated on
a held-out calibration sample $\Cset$. The constraint asks that
the proxy's expected error on the documents it would auto-accept
be at most the user's error budget $\epsilon=1-\alpha$.
Recall the score $\score_i = 2|\prob_i-\tfrac{1}{2}|$ from the
architecture above, and define the score-weighted soft error
\begin{equation}\label{eq:gated-risk}
  \widehat R_{\Cset}
  \;=\;\frac{\sum_{d_i\in\Cset} \score_i\bigl(\prob_i(1-y_i)+(1-\prob_i)y_i\bigr)}
            {\sum_{d_i\in\Cset} \score_i + \varepsilon}.
\end{equation}}
The constraint is $\widehat R_{\Cset}\le\epsilon$.

{We enforce this constraint softly with a standard
\emph{primal--dual} scheme~\cite{boyd2004convex,bertsekas1999nonlinear,cotter2019two}. The scheme attaches
a non-negative weight $\lambda\ge 0$ -- the \emph{Lagrange
multiplier}~\cite{boyd2004convex,bertsekas1999nonlinear} -- to the constraint
and adds a single penalty term
\begin{equation}\label{eq:sla-penalty}
  \mathcal{L}_{\text{sla}}
  \;=\;\lambda\cdot\max\bigl(0,\;\widehat R_{\Cset}-\epsilon\bigr)
\end{equation}
to the training loss. The training loop then alternates between
two updates: a \emph{primal} step, in which the proxy weights are
moved by one gradient descent step on the loss with $\lambda$ fixed; and a \emph{dual} step, in which $\lambda$ itself is
updated at the end of each epoch with the proxy fixed.
Intuitively, $\lambda$ is an \emph{adaptive penalty weight} -- ``how
strongly does training currently have to push the proxy to keep
$\widehat R_{\Cset}$ at or below $\epsilon$.''
The mechanics are simple. (i) When the constraint is satisfied
($\widehat R_{\Cset}\le\epsilon$), the $\max(\cdot)$ part of
\eqref{eq:sla-penalty} is zero, so $\mathcal{L}_{\text{sla}}$
vanishes from the total loss for that epoch and gradients come
only from terms (a) and (c) below; the dual step then decays $\lambda$
slightly toward $0$.
(ii) When the constraint is violated
($\widehat R_{\Cset}>\epsilon$), the $\max(\cdot)$ part becomes
positive, and the next primal step moves the proxy toward
predictions that lower $\widehat R_{\Cset}$ -- i.e., either making
the proxy more confident on documents it actually gets right, or
making it less confident on documents it gets wrong; the dual step
then raises $\lambda$ in proportion to how badly the constraint
was violated, so a \emph{persistent} violation steadily inflates
the penalty's share of the total loss until training restores
$\widehat R_{\Cset}\le\epsilon$. In practice $\lambda$ is clipped
to $[0, 300]$ and starts near $0$, so the penalty contributes to
the loss only when the proxy's auto-accept error genuinely exceeds
the user's target.}
%This constraint is not present in \methScaleDoc{}'s contrastive scheme. 
{The ablation in \S\ref{sec:experiments:rq6} shows it is the
SLA-stabilizer: removing it drops the SLA hit count from $16/17$
to $13/17$ on \dsPubMed at only a modest cost increment of
$\sim$$4\%$.
}}

{\paragraph{(c) Coverage regularizer: how do we stop the
constraint from being trivially satisfied?}
{The constraint in (b) is normalized by the proxy's score $\score$: if
the proxy reports a low score on \emph{everything} (every
$\prob_i\approx\tfrac{1}{2}$), the constraint is met by cascading
every document to the LLM. We discourage that trivial corner with
a single, lightweight term that pushes the average training-set
score up:
\begin{equation}\label{eq:cov}
  \mathcal{L}_{\text{cov}}
  \;=\;1-\tfrac{1}{|\Tset|}\sum_{d_i\in\Tset} \score_i,
\end{equation}}
%weighted at $\beta_{\text{cov}}=0.35$. 
{The ablation in \S\ref{sec:experiments:rq6} shows this term is the
cost-tightener: removing it inflates the end-to-end latency by
$\sim$$24\%$ on \dsPubMed.
}
%{raises the
%fraction of queries that miss the accuracy target by
%$\sim$$15\%$. (relative). The reason is that, with the coverage
%term removed, the constraint in (b) is no longer satisfied by
%making the proxy itself more confident; it is satisfied instead by
%routing more borderline documents to the oracle, which keeps the
%auto-accept error low but does not actually improve the proxy.}}

{\paragraph{Total loss} The total loss is ($\beta_{\text{cov}}=0.35$):
\begin{equation}\label{eq:total-loss}
  \mathcal{L}
  \;=\;\mathcal{L}_{\text{soft}}
   + \beta_{\text{cov}}\cdot\mathcal{L}_{\text{cov}}
   + \lambda\cdot\max(0,\widehat R_{\Cset}-\epsilon).
\end{equation}

{The three terms in \eqref{eq:total-loss} are not all applied to every
component of the proxy; \CE and \CB are trained with (a) only,
and the hybrid head is trained with all three, as this is the
component that produces the final probability $\prob_i$ and
therefore the score $\score_i$ the cascade thresholds. This
separation matches what each component is responsible for: (b)
and (c) shape the score $\score_i$ that determines actual
auto-label vs.\ cascade; the \CE and \CB backbones learn to align
their relevance scores with the oracle's probability and can
even be reused across different SLA targets (which we disable by default).}

%Three standard ingredients, no contrastive loss, no
%hard-negative mining. The architecture, not the training scheme,
%drives the cost reduction; the SLA-aware constraint of (b) is what
%keeps the cost reduction safe under the user's accuracy target.}

{\section{Per-Score-Range Adaptive Calibration}\label{sec:bercp}}

{The calibration step turns the proxy's score into a
threshold $\tau$: documents with score above $\tau$ are
auto-classified, the rest are sent to the oracle.
We argue that prior calibrations leave cost on the table for a single, fixable reason:
they treat all documents the same when deciding how much to trust
the labeled sample, even though that sample covers some documents
much better than others.}

\subsection{Why prior calibrations are conservative}\label{sec:bercp:why}

{To decide $\tau$, a calibration estimates
the proxy's error rate as a function of the proxy's score. 
The estimation is done on a small labeled sample
$\Cset$. \methScaleDoc{}~\cite{scaledoc} groups documents in $\Cset$ into
$64$ score ranges and uses the measured error rate inside each
range to decide $\tau$ so that the expected
accuracy on auto-accepted documents meets the target $\alpha$.}

{This works well on score ranges that contain enough 
labeled documents: the measured error rate is close to the true
error rate on all remaining docs. 
It fails on score ranges where the labeled sample is sparse: 
this may set $\tau$ too optimistic (calls the oracle less often than necessary --- an SLA violation)
or too conservative (calls the expensive oracle more often than needed).
%a single lucky or unlucky draw can shift the measured
%error rate by several percentage points, and the threshold then
%either trusts a measurement that is not really there (and calls
%the LLM less often than the target allows --- an SLA violation)
%or refuses to trust any measurement at all (and calls the LLM
%more often than the target requires --- an unnecessary cost).}

{The simplest fix is to add a worst-case safety margin
to every measured error rate, the way \methBARGAIN{}~\cite{bargain2025}
does. It replaces the measurement with a high-confidence upper bound
that holds with high probability over the labeled sample. This is
safe but expensive. The safety margin is added uniformly, including
on score ranges where the labeled sample is enough and
the measurement is already reliable. 
%The threshold ends up calling the LLM on documents the labeled sample has already pinned down,
%even though the measured error rate there is trustworthy on its
%own.
Instead, we add safety margin to score ranges
where the sample is sparse, and stay close to zero on
score ranges where the sample is enough.}

{\subsection{Per-range mix of the empirical rate and a Clopper--Pearson upper bound}\label{sec:bercp:perbin}}

We partition the labeled auto-accept set
$\Aset_\Cset(\tau)=\{d_j\in\Cset:\score_j\ge\tau\}$ into $B=20$
equal-frequency score ranges. Let $n^c_b$ be the number of
labeled documents in range $b$ and $k_b$ the number whose proxy
prediction disagrees with the oracle. \methScaleDoc uses the
empirical (measured) error rate $\hat e_b=k_b/n^c_b$. We replace
this single number with a combination of $\hat e_b$ and a
one-sided high-confidence upper bound $\mathrm{CP}_b$ on the same
quantity (computed by the standard {Clopper--Pearson
formula~\cite{clopper1934use,wilson1927probable,brown2001interval}}, %at level $\delta_b=\delta/B$,
which is the standard tool for upper-bounding a binomial rate from
a small sample; see~\cite{vovk2005algorithmic,angelopoulos2021gentle,bates2021distribution,gibbs2021adaptive,zaffran2022adaptive} for related distribution-free conformal calibrations):
\begin{equation}\label{eq:bin-bound}
  \hat u_b=(1-\lambda)\hat e_b + \lambda\,\mathrm{CP}_b,
  \qquad \lambda=0.06\text{ (default).}
\end{equation}

The mixing weight $\lambda$ is small on purpose. Here is
why $\hat u_b$ does what we wanted in the previous subsection.
\begin{itemize}[topsep=2pt,itemsep=2pt,leftmargin=1.4em]
\item When the labeled sample is plentiful in range $b$
  (large $n^c_b$), Clopper--Pearson collapses to the
  measured error rate: $\mathrm{CP}_b \approx \hat e_b$, hence
  $\hat u_b\approx\hat e_b$.
\item When the labeled sample is sparse in range $b$
  (small $n^c_b$), Clopper--Pearson widens, so
  $\mathrm{CP}_b \gg \hat e_b$, and $\hat u_b$ inflates above
  $\hat e_b$. The threshold becomes less willing to auto-accept
  documents in that range.
\end{itemize}

{In summary, $\hat u_b$ tracks the measured rate on
well-covered ranges and inflates only where the measured rate is
not reliable.}

\subsection{Threshold selection}\label{sec:bercp:certificate}

{We write $\Rset$ for the deployment pool (the unlabeled documents to classify) 
and $\Cset\subset\Rset$ for the small
labeled calibration sample. For any candidate
threshold $\tau$, $\Aset_\Rset(\tau)=\{d_i\in\Rset:\score_i\ge\tau\}$
is the subset of $\Rset$ that the proxy auto-accepts. We reuse the
score ranges from \S\ref{sec:bercp:perbin}: range $b$
($b=1,\dots,B$) is the $b$-th bucket of the
calibration scores $\{\score_j\}_{j\in\Cset}$.}

{With $\hat u_b$ in hand, the rest of the procedure is
the standard one. We project the labeled-set rates onto the
remaining unlabeled pool. At threshold $\tau$, let $n^p_b$ be the
number of pool documents the proxy auto-accepts whose score falls
in range $b$. The expected number of errors on the auto-accepted
pool is
\begin{equation}\label{eq:pool-err}
  \widehat{\mathrm{Err}}(\tau)
  \;=\;\sum_{b=1}^B n^p_b\cdot\hat u_b,
\end{equation}
and the deployed threshold is the one that calls the oracle as little
as possible subject to the accuracy target:
\begin{equation}\label{eq:bercp}
  \tau^\star=\arg\min\{|\Rset\setminus\Aset_\Rset(\tau)|
   \,:\, 1-\widehat{\mathrm{Err}}(\tau)/N\ge\alpha\}.
\end{equation}
Algorithm~\ref{alg:bercp} states the procedure.}

\begin{algorithm}[t]
\caption{Threshold selection. 
%$\Cset$ is the labeled calibration 
%sample, $\Rset$ is the deployment pool, and ``range $b$'' is the
%$b$-th equal-frequency bucket of $\score_\Cset$. 
}\label{alg:bercp}
\begin{algorithmic}[1]
\Input labeled $\Cset$, pool scores $\{\score_i\}_{i\in\Rset}$, target accuracy $\alpha$,
   number of score ranges $B$, mixing weight $\lambda$.
\Output score threshold
%{// The candidate-$\tau$ grid is the 200-quantile grid of $\score_\Cset$ plus the trivial choices $\{0,\tfrac12,1\}$.}
\For{candidate $\tau\in\mathrm{Quantiles}(\score_\Cset,200)\cup\{0,\tfrac12,1\}$}
  \State Form $\Aset_\Cset(\tau)$, partition into $B$ equal-frequency ranges.
  \For{$b=1,\dots,B$}
    \State Compute $\hat e_b$, $\mathrm{CP}_b$, $\hat u_b$ by Eq.~(\ref{eq:bin-bound}).
    \State Count $n^p_b=|\{i\in\Aset_\Rset(\tau):\score_i\in\text{range }b\}|$.
  \EndFor
  \State Compute $\widehat{\mathrm{Err}}(\tau)$ by Eq.~(\ref{eq:pool-err}).
\EndFor
\State \Return $\tau^\star$ by Eq.~(\ref{eq:bercp}).
\end{algorithmic}
\end{algorithm}

{\subsection{Why is it tighter?}\label{sec:bercp:why_tighter}

At a quick read, replacing $\hat e_b$ with $\hat u_b\ge\hat e_b$ looks \emph{more}
conservative than \methScaleDoc, not less. The key point is that
\methScaleDoc's calibration is not just the bare empirical rate
either: it smooths the per-bin counts before selecting the
threshold~\cite{scaledoc}. The smoothing is a deliberate safety
choice -- it prevents an optimistic threshold on ranges that the
labeled sample covers only sparsely -- but it increases the per-bin
error estimate \emph{uniformly} across all $B$ ranges, including
the ones the labeled sample already covers densely. In effect, 
it pays a (small) safety surcharge on every range,
whether the measurement on that range needs it or not.
In contrast, we add safety margin only where it is needed.

\subsection{Assumptions}\label{sec:bercp:assume}

{The calibration assumes (a) within each score range,
the per-document error is an independent Bernoulli draw, which is
{what the Clopper--Pearson bound~\cite{clopper1934use} needs}; and (b) the labeled sample
$\Cset$ is drawn from the same distribution as the deployment pool.
%which is the same assumption both \methScaleDoc{}'s and
%\methBARGAIN{}'s calibrations need. 
%We do not address calibration-to-pool distribution shift in this paper.
}

\section{Adaptive \methTwoPhase: Model-Free Then Proxy}\label{sec:twophase}

{Our proxy in \S\ref{sec:proxy} with the calibration in 
\S\ref{sec:bercp} is the fastest plan on non-easy predicates,
but not on \emph{every}
query: on queries where dense embeddings already group documents
the way the predicate does, \methCSV{} is able to classify all documents confidently
without many oracle calls.
However, neither method is the fastest on every case. 
We propose \methTwoPhase that composes the two so that each query
is answered by whichever phase is enough.}

\subsection{Why a composition, not a choice}\label{sec:twophase:why}

{On a typical evaluation, no single method is the
per-query winner across the whole query mix
(\S\ref{sec:experiments:rq4}). \methCSV is the fastest when
the predicate aligns with an embedding clustering --- for example,
on \dsPubMed 
{\qpubmed{3}} (``pediatric participants under 18''), a few
clusters of pediatric documents are nearly all positive and the
rest are nearly all negative, so a small per-cluster sample is
enough to label the entire cluster. Our proxy is the fastest 
when the predicate cuts across the clustering ---
for example, on \dsPubMed 
{\qpubmed{78}} (``methodological detail to assess internal
validity''), {positives and negatives are mixed inside every
cluster, so the per-cluster vote never agrees, the algorithm keeps
splitting clusters and re-sampling until it runs out of labeling
budget, and most documents end up cascaded to the oracle anyway.}}

\paragraph{The composition idea.}
{A user \emph{does not} know in advance which method is faster \emph{in the end}.
{What
we \emph{do} know is each method's predetermined \emph{must-pay} labeling cost
-- the cost the method pays before it can produce any prediction at
all, regardless of how easy or hard the query turns out to be. Our
proxy must label a fixed ${\sim}7\%$ of the corpus for training plus
${\sim}5\%$ for calibration before it can be deployed; \methCSV
must label only its first per-cluster sample (${\sim}0.5\%$ of the
corpus per cluster), and pays more only when a cluster does not
agree and has to be split. On a query where \methCSV{}'s clusters
agree quickly, its must-pay cost is well below the proxy's; on a
query where every cluster is mixed, the per-cluster samples accumulate
linearly with the corpus until \methCSV catches up with -- and
typically overshoots -- the proxy's $12\%$ budget. We use this
ordering to compose the two: run \methCSV first (its must-pay cost
is smaller), and escalate to the proxy phase only when 
the samples accumulate enough. The Phase-1 labels are
reused as the Phase-2 \emph{training} set, so the escalation adds no extra
oracle calls for training -- only the calibration set is sampled
afresh.}}

{\subsection{Procedure: model-free first, proxy only when needed}\label{sec:twophase:method}}

{\paragraph{The labeled sets.}
The deployment pool $\Rset$ is the full corpus the filter must
classify (\S\ref{sec:bercp}), and $\Cset\subset\Rset$ is the
labeled calibration sample; $\Tset\subset
\Rset$ is for the labeled training sample used by the proxy.
$\Tset$ and $\Cset$ are sampled disjointly, but both are subsets of
the same $\Rset$.}

\paragraph{Phase 1: \methCSV}
Cluster the $N$ documents into $k=4$ groups on dense embeddings. In
each round, for every cluster:
\begin{enumerate}[topsep=2pt,itemsep=1pt,leftmargin=2em]
\item Draw a sample of size $\max(\lceil 0.005\,N\rceil, 100)$;
\item Call the oracle on the sample, collecting $y$ and $\prob^\star$;
\item Vote: if the agreement on the sample is at least the vote threshold 
  $\rho_{\text{vote}}$, propagate the majority label to the
  entire cluster;
\item Otherwise, split the cluster into two via $k$-means and revisit
  it in the next round.
\end{enumerate}
{Phase~1 stops when the cumulative labeled fraction
reaches $\lambda_{\text{p1}}=0.07$ of the corpus. The vote
threshold is set to the user's deployment target,
$\rho_{\text{vote}}=\alpha$. This couples the two phases through a
single user knob. A higher target accuracy $\alpha$ makes Phase~1 more
conservative about declaring a cluster homogeneous (and so more
aggressive about reclustering), while the same $\alpha$ tightens
the Phase-2 calibration in \S\ref{sec:bercp}.}

\paragraph{Early exit.}
{If at the end of Phase~1 every cluster has agreed (which may happen
before the labeling budget $\lambda_{\text{p1}}$ is reached), the
predictions are already known and Phase~2 is skipped.}
The only
oracle cost is the Phase-1 sample.
%On \dsPubMed q3, Phase~1 alone resolves the query in $\sim$~$400$ oracle calls.

\paragraph{Phase 2 with reused training labels.}
If at least one cluster remains mixed at the end of Phase~1, we
escalate. {The Phase-1 labels become the Phase-2 training
set: every document the oracle labeled during Phase~1
($\lambda_{\text{p1}}$ of the corpus) goes into $\Tset$, with no extra
oracle calls. The calibration set $\Cset$ (0.05 of the
corpus) is sampled \emph{separately} from the pool minus
$\Tset$, by stratified sampling on the proxy score,
and does require its own oracle calls.}
\begin{enumerate}[topsep=2pt,itemsep=1pt,leftmargin=2em]
\item {Train CE, CB, and the hybrid head on
  $\Tset$ (and $\Cset$ for the primal--dual constraint).}
\item %Score all $N$ documents with the proxy (including documents in agreed
  Phase-1 clusters --- Phase~2 does \emph{not} trust propagated
  labels since the predicate is now determined non-easy).
\item Compute the score $\score_i$ for every document in the remaining
  pool $\Rset$, run \methBERCP (Algorithm~\ref{alg:bercp}) to
  select $\tau$, and cascade
  documents with $\score_i<\tau$ to the oracle.
\end{enumerate}

{\subsection{Two design choices}\label{sec:twophase:design}

The two non-obvious knobs in the procedure are the vote threshold
$\rho_{\text{vote}}$ that gates Phase-1 propagation, and the
decision \emph{not} to sample the Phase-2 calibration set $\Cset$
from the documents Phase-1 has already labeled. %Both fall out of
%choices the previous sections already made, but each is worth a
%short paragraph because neither is obvious from the procedure alone.

\paragraph{Vote threshold $=$ user target.}
The original \methCSV paper~\cite{csv2026} does not say how to set
$\rho_{\text{vote}}$ from a user accuracy target; if anything, it
explains the opposite -- the vote threshold and the realized
accuracy are not directly coupled, and tuning them by hand is part
of the cost of using \methCSV. We sidestep that tuning by taking
$\rho_{\text{vote}}=\alpha$. The reading is simple: if Phase-1
propagates a cluster's majority label only when the sample agrees
on at least an $\alpha$ fraction of its documents, then -- in
expectation -- the propagated label is wrong on at most $1-\alpha$
fraction of the cluster, i.e., the per-cluster expected error rate
is at most the user's error budget. Coupling the two settings
through one knob also makes the two phases pull in the same
direction: a stricter target accuracy makes Phase-1 more
conservative about declaring a cluster homogeneous (and so more
aggressive about reclustering), and the same $\alpha$ tightens Phase 2.

\paragraph{Calibration set is sampled separately from Phase-1.}
A natural-looking optimization is to run Phase-1 a little longer
and reuse those extra labels as the calibration set
$\Cset$, so that the transition between phases costs zero oracle
calls. However, it violates the Clopper--Pearson
assumption in \S\ref{sec:bercp:assume} that $\Cset$ is drawn from the same distribution as the deployment pool. Phase-1's sampling
is anything but uniform: it concentrates on whichever clusters
turned out to be mixed first, so a $\Cset$ taken from Phase-1's
labels is a biased view of the remaining pool, the per-bin
Clopper--Pearson bound is no longer a valid upper bound on the
pool's error rate, and the safety guarantee from \S\ref{sec:bercp}
no longer holds. We pay the cost of sampling $\Cset$ separately
(stratified on the proxy score over the pool minus $\Tset$) so
that the calibration set is representative of the documents the
threshold will actually act on.}

\section{BER as Difficulty Compass and Lower Bound}\label{sec:headroom}

{We propose two uses of the oracle's own per-document
confidence. \emph{(i)~A difficulty compass.} The same per-document confidence
that we use as a soft training label in \S\ref{sec:proxy} is also
a method-independent measure of how hard a query is: it predicts,
without running any method, which one will be
fastest on that query.
\emph{(ii)~A lower bound on cascade cost.} The same per-document
confidence places a hard floor on how few documents \emph{any}
cascade plan can send to the oracle while still meeting the
accuracy target, regardless of how smart the proxy is. We use the
floor as a benchmark for headroom and as a research target for
future work; we do not claim it is attainable.}

\subsection{Bayes error: the oracle's own uncertainty}\label{sec:headroom:def}

{For document $d_i$, let $\prob^\star_i\in[0,1]$ be the
oracle's predicted probability that the document satisfies the
predicate. The \emph{per-document Bayes error}~\cite{fukunaga1990introduction,northcutt2021confident} is
$\eta_i=\min(\prob^\star_i,1-\prob^\star_i)$: the probability that
the oracle's own hard label is wrong if we were to re-sample it
from the oracle. A document the oracle is sure about
($\prob^\star_i$ close to $0$ or $1$) has $\eta_i$ close to $0$;
a document the oracle is unsure about ($\prob^\star_i$ close to
$\tfrac{1}{2}$) has $\eta_i$ close to $\tfrac{1}{2}$. So the ceiling is $\tfrac{1}{2}$.}

{The \emph{query BER} $\overline\BER_q=\tfrac{1}{N}\sum_i\eta_i$
averages this over the corpus. It is a property of the query and
the oracle; it does not depend on which proxy or which calibration
the user picks. It is also free to measure: the oracle returns
$\prob^\star_i$ together with the hard label on every call, so
$\overline\BER_q$ can be estimated from any pilot sample at no
extra oracle cost.}

\subsection{Why query BER predicts which family wins}\label{sec:headroom:compass}

{The two prior families behave very differently as
$\overline\BER_q$ grows.
\methCSV labels a small per-cluster sample and propagates the
majority label to the whole cluster. When $\overline\BER_q$ is
small, most documents are confidently in one class, so most
clusters are homogeneous and propagation gets the answer for free.
As $\overline\BER_q$ grows, more documents fall near the predicate
boundary, more clusters become mixed, and \methCSV must split them
and label larger samples; the cost grows quickly.
An online proxy (\methScaleDoc or our \methBERCP) pays a fixed
training and calibration cost regardless of $\overline\BER_q$, and
a deployment cost that grows much more slowly with
$\overline\BER_q$ because the proxy can resolve documents the
cluster vote could not.
The two cost curves cross at a query-specific BER (\S\ref{sec:experiments:rq5}): below the crossing,
\methCSV is the fastest plan; above it, our proxy is the fastest. 
\methTwoPhase tracks the faster of the two on each query by
running \methCSV first and escalating only when \methCSV is not
enough.}

\subsection{A lower bound on deployed cascade calls}\label{sec:headroom:lb}

{The Bayes error $\eta_i$ is a lower bound on the
probability that \emph{any} classifier, on any representation,
makes a mistake on $d_i$~\cite{fukunaga1990introduction}, because
the oracle is the most accurate label we have. Suppose a cascade
plan auto-classifies a subset $\Aset$ of the pool and sends the
rest to the oracle. The expected number of auto-classification
errors is at least $\sum_{d_i\in\Aset}\eta_i$. To meet the corpus
error budget $(1-\alpha)N$, the plan must keep this sum below the
budget, which means it can only auto-classify documents whose
$\eta_i$ values are small enough to fit inside the budget when
summed.}

\paragraph{\methBERLB}
{The cheapest set $\Aset$ that fits inside the budget
is found by sorting documents in increasing order of $\eta_i$ and
including them one by one until the budget is full:}

\begin{definition}[\methBERLB]\label{def:berlb}
{Sort documents by ascending Bayes error
$\eta_{(1)}\le\eta_{(2)}\le\cdots\le\eta_{(N)}$, and let
$k^\star$ be the largest $k$ with
$\sum_{i=1}^k\eta_{(i)}\le(1-\alpha)N$. Then the minimum number
of cascade oracle calls any cascade plan can make at the target
accuracy $\alpha$ is at least $N-k^\star$.}
\end{definition}

{\methBERLB does not count the oracle calls needed to
\emph{learn} the per-document $\eta_i$ (it assumes them as given),
so it is a strict lower bound on the cost of deployed cascade
calls, not a target a real method can reach. We use it to
quantify headroom: the gap between the cheapest deployable plan
and \methBERLB tells us how much further any future cascade method
could possibly cut the cost.}

\subsection{Headroom}\label{sec:headroom:headroom}

{Our deployable methods sit closer to
\methBERLB than the baselines, but the floor itself is still
{$\sim$$4$--$20\times$} below our cost on the three corpora
(\S\ref{sec:experiments:rq1}).
Two costs explain the remaining
gap: \emph{(i)} the oracle calls needed to learn the per-document
$\eta_i$ values, which \methBERLB does not pay; and \emph{(ii)}
the safety margins needed to certify the accuracy target under a
finite labeled sample. Closing the gap is a concrete research
target for future cascade plans.}

\paragraph{Summary.}
{Oracle Bayes error gives the user, for free, both a
prediction of which cascade family will be cheapest on a query
and a hard lower bound on the cost of any cascade family on that
query. Neither use has been isolated in prior semantic-filter
work; we use both inside \methTwoPhase and as headroom in the
evaluation.}

\section{Experiments}\label{sec:experiments}

{This section asks whether the four contributions of \S\ref{sec:intro}
other than the unified framework (C1)---the revisited online proxy
(C2, \S\ref{sec:proxy}), the tighter calibration (C3,
\S\ref{sec:bercp}), the adaptive two-phase method (C4,
\S\ref{sec:twophase}), and the BER compass and lower bound (C5,
\S\ref{sec:headroom})---translate into measurable wins for the
semantic-filter operator: lower end-to-end latency at the user's
accuracy target SLA.}

We organize the evaluation around six research questions.
RQ1--RQ3 establish empirical efficacy at the user's target accuracy,
under a sweep of that target, and at a per-segment level of cost
breakdown; RQ4--RQ5 explain \emph{which} queries our methods win
and why; RQ6 isolates which ingredient of the proxy and
calibration contributions binds via controlled ablations.

\begin{itemize}[topsep=2pt,itemsep=2pt,leftmargin=2em]
\item \textbf{RQ1 (efficiency at the SLA).} At target accuracy
  $\alpha\!=\!0.9$, how much do \methBERCP and \methTwoPhase reduce
  mean end-to-end latency and oracle-call count relative to
  \methScaleDoc / \methCSV / \methBARGAIN, and how close do they get
  to the \methBERLB lower bound?
\item \textbf{RQ2 (target-accuracy robustness).} As the user
  tightens $\alpha$ from $0.70$ to $0.95$, do our methods' cost
  advantages widen, hold, or collapse?
\item \textbf{RQ3 (cost decomposition).} On a per-query basis, where
  does the time go---proxy training/scoring, sample labeling for voting, training set
  labeling, calibration set labeling, or oracle cascade?
\item \textbf{RQ4 (per-query competitiveness).} The mean numbers
  hide which plan is the fastest per query. Does a single method track the
  per-query lower envelope?
\item \textbf{RQ5 (BER as compass).} Does the query's Bayes error
  rate $\overline\BER_q$, computed from the oracle's free
  logprobs, predict which family wins on that query?
{\item \textbf{RQ6 (ablations).} Inside the
proxy, is the binding ingredient the \emph{architecture}, the
\emph{training loss}, the \emph{primal--dual} update, or the
\emph{coverage regularizer}? Inside the calibration, where do our
single-knob per-bin blend, \methScaleDoc's histogram band, and a
{naive empirical calibration} sit relative to the non-deployable
\emph{omniscient bound} that knows every label in advance?}
\end{itemize}

\subsection{Setup}\label{sec:experiments:setup}

\paragraph{Datasets and queries.}
We use the three corpora used in \cite{scaledoc}, each with $N\!\approx\!10\,000$ documents, along with $20$ queries we generated:
\textbf{\dsPubMed} (biomedical abstracts, mean $510$ prompt tokens),
\textbf{\dsBigPatent} (patent specifications, $233$ tokens) and
\textbf{\dsGovReport} (government reports, $718$ tokens).
%The three corpora are chosen to span short/medium/long prompts, 
%so per-oracle-call latency varies by $\sim$$3\times$
%across them, exposing methods that win or lose because of the
%\emph{number} of oracle calls rather than their cost.
Document features are \NVEmbed embeddings (4096-D) \cite{lee2025nvembed,wang2022e5,muennighoff2023mteb} shared across
methods. The oracle is \oracleLLM~\cite{llama3,touvron2023llama} run on two A100 80GB GPUs.

\paragraph{Methods compared.}
\begin{itemize}[topsep=2pt,itemsep=2pt,leftmargin=2em]
\item \textbf{\methCSV}~\cite{csv2026}: $k$-means $k=4$, sample
  {$\max(\lceil 0.005\,N\rceil, 100)$ documents}
  per cluster,
  {vote threshold $\rho_{\text{vote}}=\alpha$},
  same as described in \S\ref{sec:twophase:method}.
\item \textbf{\methBARGAIN}~\cite{bargain2025}: \proxyLLM is used as a prebuilt proxy.
\item \textbf{\methDefault}~\cite{scaledoc}:
  bi-encoder over frozen \NVEmbed trained with multi-stage contrastive learning, 
  $7\%$ train / $5\%$ calibration split.
\item \textbf{\methBERCP (ours)}: Our proxy in \S\ref{sec:proxy} and calibration in \S\ref{sec:bercp}. Same $7\%$/$5\%$ splits as \methDefault.
\item \textbf{\methTwoPhase (ours)}: \methCSV $\to$ \methBERCP, with
  Phase-1 labels reused for Phase-2 training (\S\ref{sec:twophase}).
\item \textbf{\methBERLB} (Def.~\ref{def:berlb}): non-deployable lower
  bound based on the oracle's per-document Bayes errors.
\end{itemize}

\paragraph{Primary metrics.}
We use four metrics:
\emph{(i)~end-to-end latency},
\emph{(ii)~oracle-call count} (the dominant cost),
\emph{(iii)~SLA hit count} (number of queries whose
realized corpus accuracy $\acc_q$ meets the target $\alpha$), and
\emph{(iv)~SLA-violation magnitude} ($\sum_{q}\max(0,\alpha-\acc_q)$ summed
over queries; lower is better, $0$ means no query violated
the SLA). The hit count and the violation magnitude together
distinguish a calibration, e.g., that misses on $4/20$ queries by a hair
from one that misses on $1/20$ by a wide margin.

{\paragraph{Hyperparameters.}
Unless otherwise stated, \methBERCP uses $B=20$ bins,
$\lambda_{\text{cp}}=0.06$ as the weight for the CP blend in our calibration. 
\methTwoPhase uses vote threshold
$\rho_{\text{vote}}=\alpha$ as \methCSV and training sample ratio $\lambda_{\text{p1}}=0.07$ as \methScaleDoc; the proxy trains the CE for $60$ epochs, CB for $15$, hybrid
head for $120$.}

\subsection{RQ1: Efficiency at the SLA}\label{sec:experiments:rq1}

Table~\ref{tab:e2e} summarizes mean end-to-end
latency, mean oracle-call count, SLA hit count, and SLA-violation
magnitude on three corpora at the standard target $\alpha=0.9$.
The non-deployable \methBERLB row quantifies remaining headroom.
Three findings stand out.

\paragraph{\methBERCP cuts E2E and oracle calls by
roughly $2\times$.}
On \dsPubMed \methBERCP is $2.03\times$ faster than \methDefault
($264.6$\,s vs.\ $536.8$\,s); on \dsGovReport it is $1.85\times$
faster ($412.7$ vs.\ $765.0$); on \dsBigPatent it is $1.87\times$
faster ($116.1$ vs.\ $217.3$). The reduction is driven almost
entirely by the cascade segment of the cost: the mean oracle-call
count falls from $3$,$468$ to $1$,$890$ on \dsPubMed
(a $\sim$$45\%$ cut), from $3$,$694$ to $2$,$141$ on \dsGovReport
($\sim$$42\%$), and from $2$,$292$ to $1$,$672$ on \dsBigPatent
($\sim$$27\%$). RQ3 traces this reduction to the per-query
breakdown.

\paragraph{\methTwoPhase is the strongest SLA-compliant plan.}
{\methTwoPhase pays a small E2E increment over \methBERCP on \dsPubMed ($8.6\%$) 
and is in fact \emph{faster} than
\methBERCP on \dsGovReport ($4.0\%$) and
\dsBigPatent ($9.4\%$).
In exchange, \methTwoPhase{} delivers $19/20$ SLA hits on every
corpus, which is the highest among the deployable plans on
\dsBigPatent and ties or trails on the others.
The SLA-violation magnitude tells the more important story.
\methTwoPhase{}'s violation is at most $0.008$ on every corpus
($0.0071$, $0.0051$, $0.0081$), roughly $7\times$ smaller than
\methBERCP{}'s $0.0551$ on \dsPubMed and below the
$0.012$--$0.023$ range of \methDefault.
The efficiency of \methTwoPhase{} comes from bypassing Phase~2 on
queries that Phase~1 alone resolves -- $3/20$ on \dsPubMed, $0/20$
on \dsGovReport, $5/20$ on \dsBigPatent.
%-- saving the calibration set ($\sim$$5\%$ of the corpus) and the
%Phase-2 cascade cost on those queries. 
The improved SLA hit count
comes from setting the Phase-1 vote threshold to the user target
$\alpha$ (\S\ref{sec:twophase:design}), which keeps Phase-1
propagation conservative.}

%The premium comes from
%Phase~2 firing on $16/20$, $16/20$, and $9/20$ queries on \dsPubMed
%/ \dsGovReport / \dsBigPatent respectively; on the bypassed
%queries Phase~1 alone resolves the predicate at the cost of a
%single $\sim$$7\%$ labelled sample.

\paragraph{Headroom: the \methBERLB floor is still
$4$--$20\times$ ahead.}
The \methBERLB row of Table~\ref{tab:e2e} reports the minimum
oracle-call count and the corresponding E2E that \emph{any}
proxy-based plan could attain at $\alpha\!=\!0.9$
(Def.~\ref{def:berlb}). The deployable-to-floor gap is
$\sim\!20.0\times$ on \dsPubMed ($287.4$\,s vs.\ $14.4$\,s),
$\sim\!7.5\times$ on \dsGovReport ($396.1$ vs.\ $53.1$), and
$\sim\!3.9\times$ on \dsBigPatent ($105.2$ vs.\ $26.7$). The gap is
largest where the BER distribution is most skewed toward zero
(\dsPubMed) and smallest where \methCSV's cluster-vote already
exploits much of the easy regime (\dsBigPatent), consistent with
the floor analysis of \S\ref{sec:headroom}.

\begin{table*}[t]
\centering\small
\setlength{\tabcolsep}{3.5pt}\renewcommand{\arraystretch}{1.15}
\begin{tabular}{@{}lrrcrrrcrrrcr@{}}
\toprule
& \multicolumn{4}{c}{\textbf{\dsPubMed}} & \multicolumn{4}{c}{\textbf{\dsGovReport}} & \multicolumn{4}{c}{\textbf{\dsBigPatent}}\\
\cmidrule(lr){2-5}\cmidrule(lr){6-9}\cmidrule(lr){10-13}
\textbf{Method}
 & E2E (s) & Oracle & acc$\ge\!0.9$ & SLA viol.
 & E2E (s) & Oracle & acc$\ge\!0.9$ & SLA viol.
 & E2E (s) & Oracle & acc$\ge\!0.9$ & SLA viol. \\
\midrule
\methCSV~\cite{csv2026}
 &  914.8 & 6933 & 19/20 & 0.0039
 & 1250.6 & 6737 & 19/20 & 0.0107
 &  255.5 & 4239 & 20/20 & 0.0000 \\
\methBARGAIN~\cite{bargain2025}
 &  579.8 & 3144 & 20/20 & 0.0000
 &  639.4 & 2194 & 20/20 & 0.0000
 &  209.1 & 2218 & 17/20 & 0.0635 \\
\methDefault~\cite{scaledoc}
 &  536.8 & 3468 & 17/20 & 0.0231
 &  765.0 & 3694 & 19/20 & 0.0123
 &  217.3 & 2292 & 17/20 & 0.0192 \\
\textbf{\methBERCP (ours)}
 & \textbf{264.6} & \textbf{1890} & \textbf{16/20} & \textbf{0.0551}
 & \textbf{412.7} & \textbf{2141} & \textbf{18/20} & \textbf{0.0065}
 & \textbf{116.1} & \textbf{1672} & \textbf{18/20} & \textbf{0.0171} \\
\textbf{\methTwoPhase (ours)}
 & \textbf{287.4} & \textbf{2080} & \textbf{19/20} & \textbf{0.0071}
 & \textbf{396.1} & \textbf{2052} & \textbf{19/20} & \textbf{0.0051}
 & \textbf{105.2} & \textbf{1555} & \textbf{19/20} & \textbf{0.0081} \\
\midrule
\methBERLB (Def.~\ref{def:berlb})
 &   14.4 &    109 & 20/20 & 0.0000
 &   53.1 &    286 & 20/20 & 0.0000
 &   26.7 &    444 & 20/20 & 0.0000 \\
\bottomrule
\end{tabular}
\caption{
{\textbf{End-to-end comparison at target accuracy $\alpha\!=\!0.9$ (SLA).}
Per dataset: mean end-to-end latency (s), mean \emph{total}
oracle-LLM call count per query, number of queries (out of
$20$) meeting the SLA, and the SLA-violation magnitude
$\sum_q \max(0, \alpha - \acc_q)$ (lower is better; $0$ means no
query violated the SLA). \methBERLB is the non-deployable greedy
lower bound of Def.~\ref{def:berlb}.}}
\label{tab:e2e}
\end{table*}

\subsection{RQ2: Target-Accuracy Robustness}\label{sec:experiments:rq2}

The $\alpha\!=\!0.9$ comparison of RQ1 is a single slice of a curve.
A calibration that wins at $\alpha\!=\!0.9$ can lose at $\alpha\!=\!0.95$
if its bound is loose on small samples; a method that wins at
$\alpha\!=\!0.85$ can become dominated when the target tightens.
RQ2 sweeps $\alpha\in[0.70, 0.95]$ and asks whether our advantage
\emph{widens}, holds, or collapses as the target tightens
(Fig.~\ref{fig:alpha_sweep}).

The pattern is consistent across corpora.
\methTwoPhase and \methBERCP form the cheapest, steepest pair at
every $\alpha$ on every corpus: they are furthest to the left
(cheapest at a given $\alpha$) and most vertical (cheapest delta
as $\alpha$ tightens), because the \methBERCP bound only inflates
the grey zone where the calibration sample is sparse.
\methDefault and \methBARGAIN trail by a roughly constant horizontal
offset across the entire $\alpha$ range, indicating that their
calibrations do not adapt to the tightening budget.
\methCSV is competitive at low $\alpha$ where Phase-1 alone
suffices, then degrades sharply as $\alpha$ tightens because the
cluster-vote residual stops fitting the budget.
At $\alpha\!=\!0.95$ all per-query proxy methods converge toward a
common wall: the global BER floor of 
{Def.~\ref{def:berlb}} leaves no
headroom on the high-BER queries, an empirical manifestation of
the \methBERLB lower bound.

\begin{figure}[t]
\centering
\begin{subfigure}[t]{\linewidth}\centering
  \includegraphics[width=0.9\linewidth]{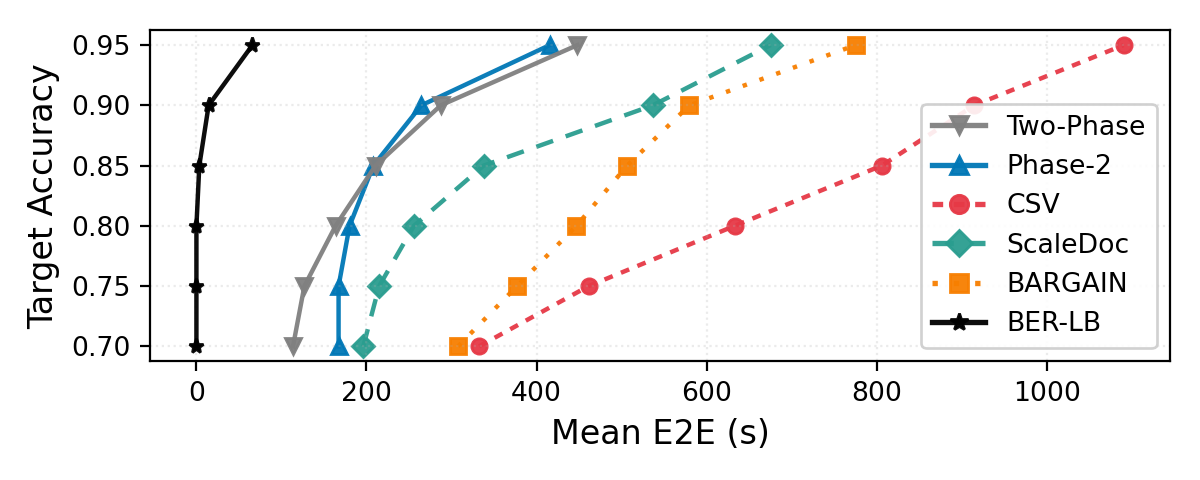}
  \caption{\dsPubMed.}\label{fig:alpha_sweep_pubmed}
\end{subfigure}
\vspace{4pt}
\begin{subfigure}[t]{\linewidth}\centering
  \includegraphics[width=0.9\linewidth]{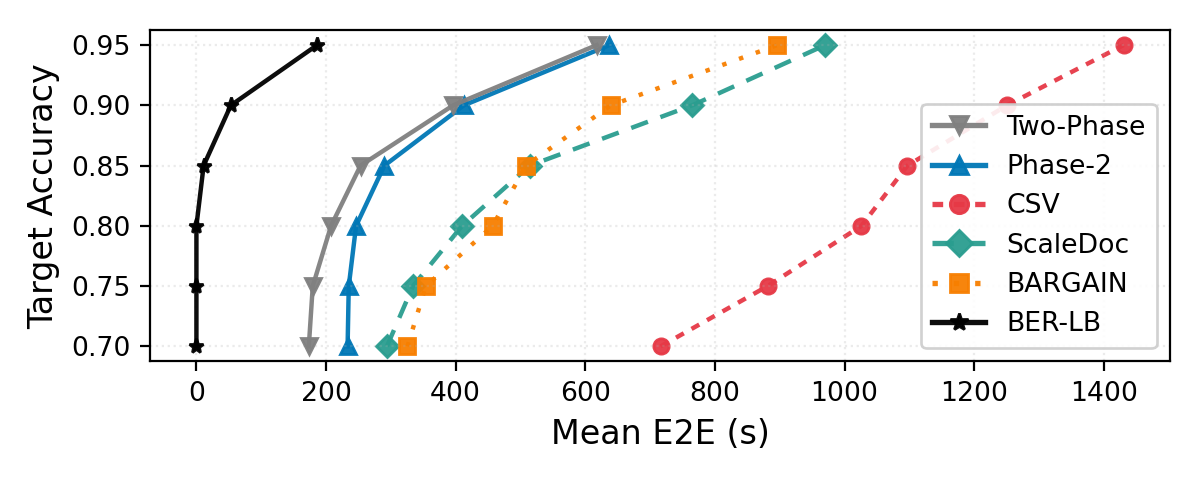}
  \caption{\dsGovReport.}\label{fig:alpha_sweep_gov}
\end{subfigure}
\vspace{4pt}
\begin{subfigure}[t]{\linewidth}\centering
  \includegraphics[width=0.9\linewidth]{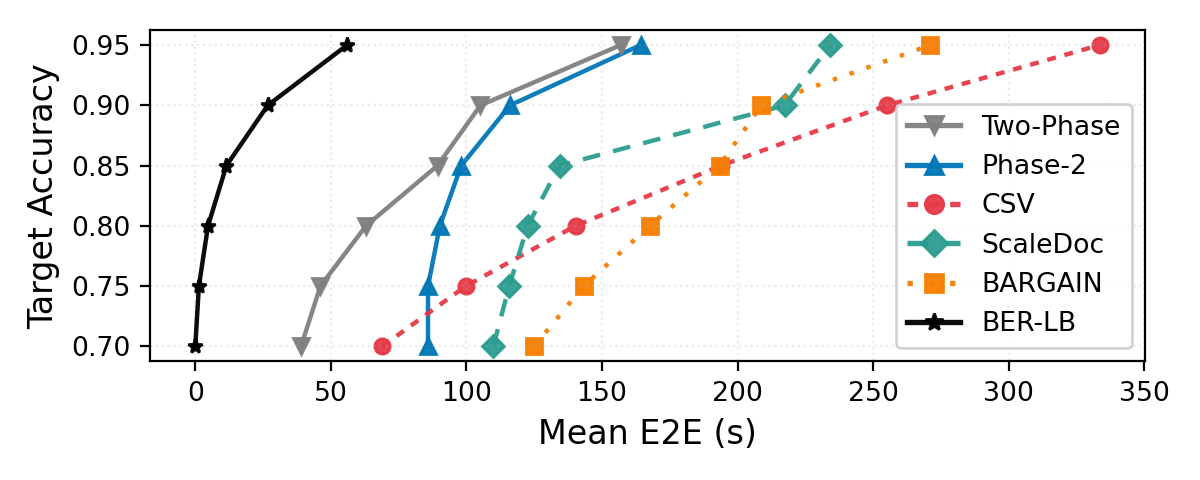}
  \caption{\dsBigPatent.}\label{fig:alpha_sweep_bigpatent}
\end{subfigure}
\caption{
{\textbf{Target accuracy vs.\ end-to-end cost.}
Curves further to the \emph{left} are cheaper at a given target;
curves steeper \emph{upward} hold their cost as the target
tightens. \methTwoPhase and \methBERCP form the cheapest,
steepest pair; \methDefault and \methBARGAIN trail by a roughly
constant offset; \methCSV is competitive only at low $\alpha$.
The dashed \methBERLB curve is the non-deployable greedy lower
bound (Def.~\ref{def:berlb}) at each $\alpha$.}}
\label{fig:alpha_sweep}
\end{figure}

\subsection{RQ3: Per-Query Cost Decomposition}\label{sec:experiments:rq3}

RQ1 and RQ2 reported aggregate cost; RQ3 asks where the time
actually \emph{goes} on each individual query.
Fig.~\ref{fig:breakdown} decomposes the per-query latency at
$\alpha\!=\!0.9$ into the five cost segments of the unified
template: proxy training/scoring, Phase-1 sample labeling,
training-set labeling, calibration-set labeling, and cascade.
Three patterns hold across all $60$ query slots.

\emph{(i)~Cascade dominates.} The cross-hatched cascade
segment is by far the largest contributor on every method that
deploys a proxy. Improvements in mean E2E come almost entirely
from shrinking this segment; proxy training time is a flat
$\sim$$15$\,s for \methBERCP and \methTwoPhase, regardless of
corpus or query.
\emph{(ii)~\methBERCP and \methTwoPhase shrink the cascade
consistently.} The cascade reduction is visible per query, not
just on average, which rules out the possibility that the mean
gain is driven by a few easy outliers.
\emph{(iii)~\methTwoPhase additionally removes the training
labeling segment on Phase-1-resolved queries.} The bars for
short-circuit queries (left side of each panel) show only the
Phase-1 sample bar; the train and calibration bars vanish because
Phase-2 is bypassed.
Absolute scale grows with prompt length (\dsGovReport's prompts
are $\sim$$3\times$ longer than \dsBigPatent's), but the relative
ordering of methods is unchanged across panels. 
%, which we take as
%evidence that the gains do not depend on prompt length.

\begin{figure*}[t]
\centering

% Shared legend on top
\includegraphics[width=0.65\linewidth]{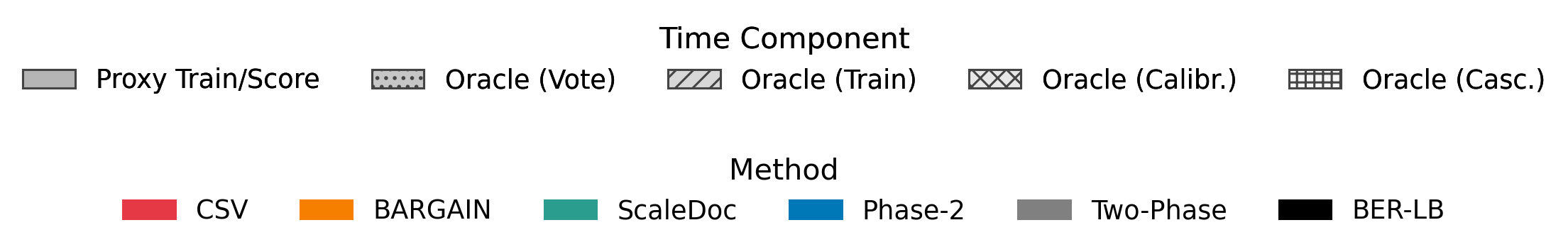}
\vspace{2pt}

% Panel 1: PubMed
\begin{subfigure}[t]{\linewidth}\centering
  \includegraphics[width=0.9\linewidth]{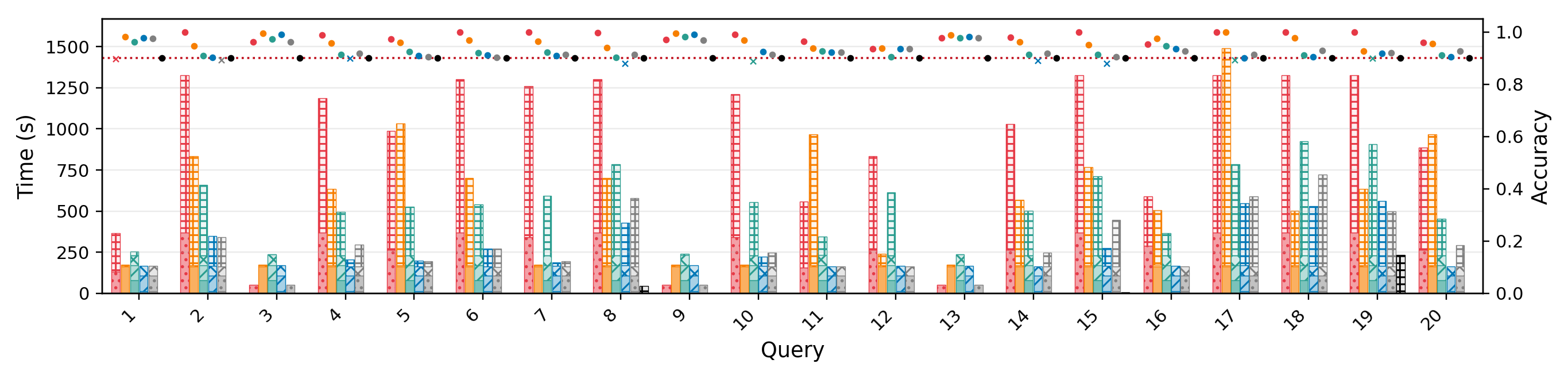}
  \caption{\dsPubMed.}
  \label{fig:breakdown_pubmed}
\end{subfigure}

\vspace{4pt}

% Panel 2: GovReport
\begin{subfigure}[t]{\linewidth}\centering
  \includegraphics[width=0.9\linewidth]{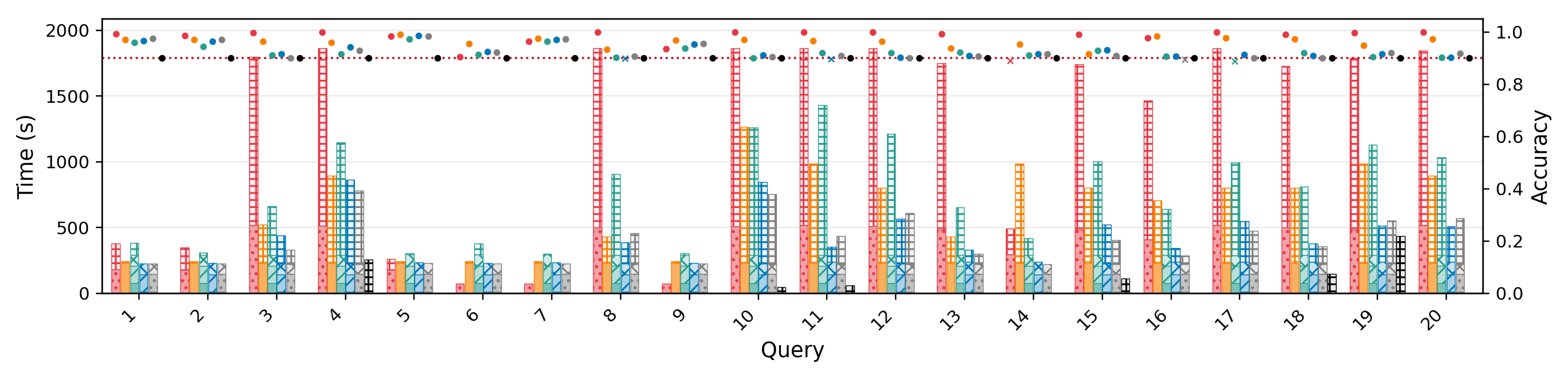}
  \caption{\dsGovReport.}
  \label{fig:breakdown_gov}
\end{subfigure}

\vspace{4pt}

% Panel 3: BigPatent
\begin{subfigure}[t]{\linewidth}\centering
  \includegraphics[width=0.9\linewidth]{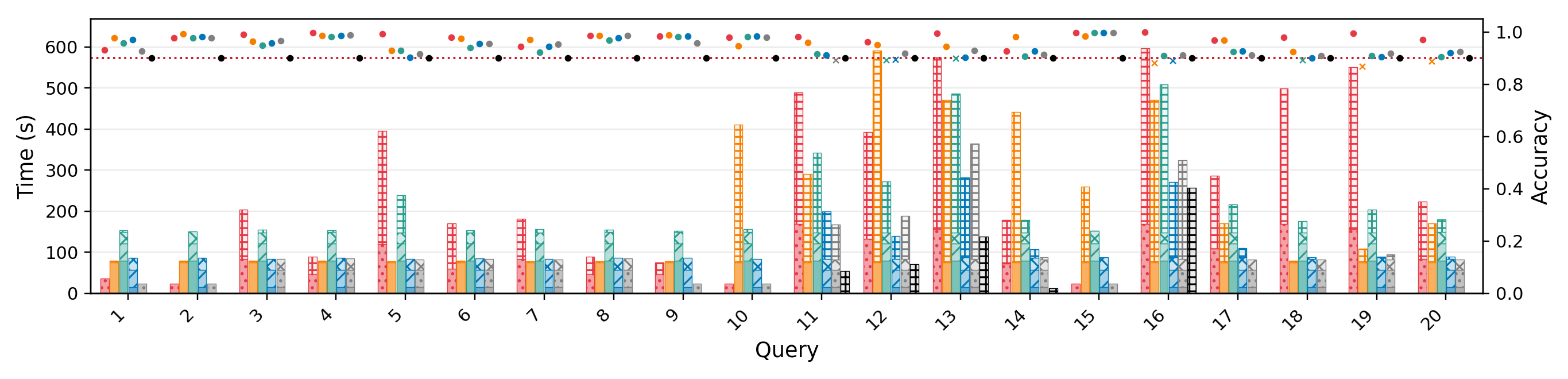}
  \caption{\dsBigPatent.}
  \label{fig:breakdown_bigpatent}
\end{subfigure}

\caption{\textbf{Per-query end-to-end time breakdown at
$\alpha\!=\!0.9$.} Bars stack proxy time and oracle-LLM time across
five segments: proxy train/score, Phase-1 sample labeling, training-set
labeling, calibration labeling, and cascade. 
The cross-hatched cascade segment dominates everywhere; \methBERCP and \methTwoPhase
shrink it on essentially every query; \methTwoPhase additionally
removes the training-labeling segment on Phase-1-resolved queries.
Right-axis markers indicate per-query accuracy:
$\circ$ for $\acc_q\ge 0.9$, $\times$ for $\acc_q<0.9$.}
\label{fig:breakdown}
\end{figure*}

\subsection{RQ4: Per-Query Competitiveness}\label{sec:experiments:rq4}

The aggregates of Table~\ref{tab:e2e} hide that the cheapest plan changes per
query. A plan that is $2\times$ better \emph{on average} but
$5\times$ worse on a small fraction of queries can still hurt the
user. The right diagnostic is a \emph{lower-envelope} view: per
query, take the cheapest deployable plan; sort the queries by this
per-query minimum; then plot each method's per-query cost against
the sorted axis. A method that ``tracks the envelope'' across the
whole axis is the plan a single-system optimizer should reach for
as a default.

\begin{figure}[t]
\centering
\includegraphics[width=0.9\columnwidth]{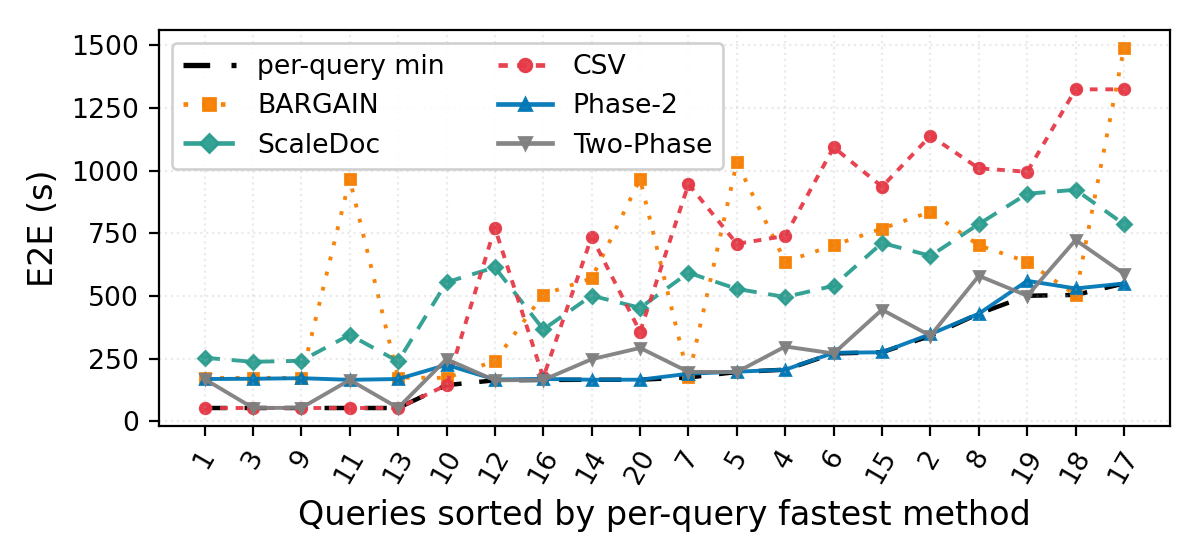}
\caption{{\textbf{Per-query lower-envelope on \dsPubMed at
$\alpha\!=\!0.9$.} Queries are sorted along the $x$-axis by the
per-query cheapest deployable plan (the dashed envelope);
$y$-axis is end-to-end time. \methTwoPhase (gray) tracks the
envelope across the whole axis; \methCSV (red) wins outright on
the leftmost $\sim$$6$ queries and then explodes by $5$--$25\times$
on the rest; \methDefault (teal) and \methBARGAIN (orange) sit far
above the envelope almost everywhere.}}
\label{fig:lowerenvelope}
\end{figure}

Fig.~\ref{fig:lowerenvelope} confirms the prediction on
\dsPubMed. \methTwoPhase tracks the envelope across all $20$
queries, being competitive to \methCSV on the Phase-1-resolved low-BER queries
and competitive to \methBERCP on the rest. \methCSV is the per-query
winner where it wins, but its loss on the right half of the axis
($5$--$25\times$ over the envelope) is what a deployer pays for
picking it as a single-system default. \methDefault and \methBARGAIN
sit far above the envelope almost everywhere, consistent with the
mean numbers of RQ1. The picture is qualitatively the same on
\dsGovReport and \dsBigPatent.

\subsection{RQ5: BER as a Compass}\label{sec:experiments:rq5}

Section~\ref{sec:headroom:compass} predicts that the per-query
ordering of methods seen in Fig.~\ref{fig:lowerenvelope} is
explained by the query's Bayes error rate $\overline\BER_q$
computed from the oracle's free logprobs. We test this directly.
For every query we record $(\overline\BER_q,\text{winner})$, where
the winner is the deployable method with the smallest E2E, and fit
a logistic regression of
$P(\methCSV{}\text{ wins} \mid \overline\BER_q)$. The crossover
BER (where the fit crosses $0.5$) is the BER above which our proxy
family takes over; AUC quantifies how cleanly $\overline\BER_q$
alone separates the regimes.

\begin{figure}[t]
\centering
\includegraphics[width=\columnwidth]{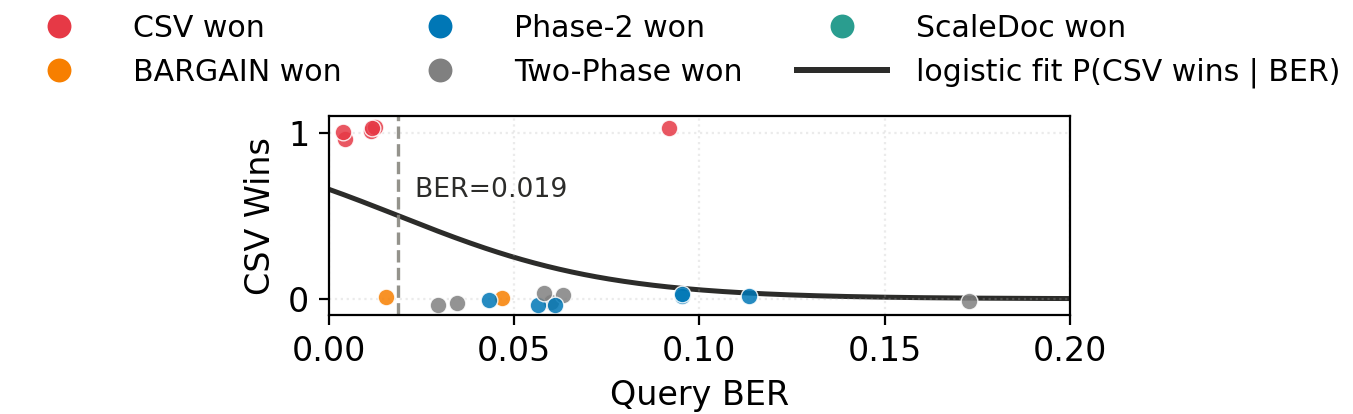}
\caption{\textbf{BER as a compass for which family wins, \dsPubMed.}
Each point is one of the $20$ queries plotted at its query BER
$\overline\BER_q$; the point's color marks the deployable method
that achieved the smallest E2E on that query. The S-curve is a
logistic fit of $P(\methCSV{}\text{ wins}\mid\overline\BER_q)$;
the dashed vertical line is the crossover BER at which the fit
equals $0.5$. The fitted crossovers are $0.020$ on \dsPubMed,
$0.040$ on \dsGovReport, $0.040$ on \dsBigPatent;
AUC of the BER-only predictor is $0.881$, $0.810$, $0.859$
respectively.}
\label{fig:compass}
\end{figure}

Two observations:
\emph{(i)~AUC $\ge 0.81$ on every corpus.}
$\overline\BER_q$ alone is a strong predictor of whether \methCSV
is the cheapest plan. The remaining gap to $1.0$ comes from a
handful of ``\methCSV beats its compass'' edge cases (\dsPubMed
{\qpubmed{22}} at $\overline\BER=0.090$ and \dsBigPatent \qbp{41} at $0.110$)
where clustering happens to align with the predicate despite high
oracle noise.
\emph{(ii)~The crossover BER is consistent within $2\times$ across
corpora} ($0.020$ on \dsPubMed, $0.040$ on \dsGovReport and
\dsBigPatent). The crossover is not a corpus-specific artifact;
it is a property of the operator. It also justifies the design of
\methTwoPhase: 
{the Phase-1 cluster-vote agreement is itself a cheap, in-pipeline signal that the query is in the low-BER regime where \methCSV is the cheaper plan with enough confidence: a query whose clusters all agree at the user target $\alpha$ is effectively a low-BER query and the rest of the budget can be saved by skipping Phase~2. \methTwoPhase therefore does not need to estimate $\overline\BER_q$ or consult any router; the Phase-1 step is the per-query plan selector.}

\subsection{RQ6: Ingredient Ablations}\label{sec:experiments:rq6}

RQ6 isolates which ingredient binds inside the two ablatable
contributions of our method: the proxy (RQ6a, four ingredients)
and the calibration (RQ6b, three ingredients).
All ablations are run on \dsPubMed; both tables restrict to the
$17/20$ queries where \methTwoPhase fires Phase~2 so that all rows
share the same train/cal split. Each table holds one contribution
fixed and varies the other.

\paragraph{RQ6a: Proxy ingredients.}
The proxy contribution has four ingredients: the
\emph{architecture} (CE+CB+hybrid head vs.\
\methScaleDoc{}'s bi-encoder), the \emph{training loss} on
CE+CB (soft-label BCE vs.\ hard-label BCE vs.\ contrastive),
the \emph{primal--dual} (PD) update that enforces the SLA
constraint during training of the hybrid head, and the
\emph{coverage penalty} (cov) that pushes the selection gate to
cover as many documents as possible. Table~\ref{tab:proxy_ablation}
sweeps each ingredient with the calibration held fixed at the
full \methBERCP.

{The ablation shows that the four ingredients have different binding strengths.}
\emph{(i)~The architecture is the dominant ingredient.}
Holding training fixed at soft-BCE + PD + cov, swapping our
CE+CB+hybrid for a bi-encoder over frozen embeddings regresses
the E2E from $328.7$\,s to $415.4$\,s and drops the SLA hits from
$16/17$ to $13/17$. The architecture is the binding constraint, as
predicted by \S\ref{sec:proxy:diag}: cosine over dense embeddings
fuses token-level evidence away, and any training scheme we evaluate cannot recover
what the representation has discarded.
\emph{(ii)~The training loss for the backbones matters at the
margin.} Holding the architecture at our CE+CB+hybrid, swapping
soft-BCE for contrastive on CE/CB is within noise on E2E
($329.6$\,s vs.\ $328.7$\,s) but drops $2$ SLA hits; swapping for
hard-BCE regresses E2E by $\sim$$8\%$ and drops $4$ SLA hits
because the proxy becomes over-confident near the cascade boundary.
\emph{(iii)~PD and cov each contribute, in different ways.}
{Removing the coverage regularizer inflates the cascade to $406.4$\,s but holds $15/17$ SLA hits;
removing PD gives a tighter $340.5$\,s but drops to
$13/17$ SLA hits. PD is the SLA-stabilizer; cov is the
cost-tightener.}
\emph{(iv)~Reference.} \methDefault's bi-encoder + contrastive
without PD/cov sits at $543.1$\,s and $13/17$, which is the gap
we attribute to the combined proxy contribution.

\paragraph{RQ6b: Calibration.}
{The calibration step holds the proxy fixed at our full
CE+CB+hybrid + soft-BCE + PD + cov and asks which procedure
turns the proxy's score into a deployable SLA.
Table~\ref{tab:cal_ablation} compares four calibrations:
{\emph{Naive empirical}} (the bare empirical error rate on
$\Cset$ with no conservative refinement); \methScaleDoc{}'s
smoothed histogram band; ours (the single-knob per-bin
Clopper--Pearson blend of \S\ref{sec:bercp:perbin}); and a
non-deployable \emph{omniscient bound} that assumes the proxy's
score \emph{and} every document's true label are known in advance
and picks the threshold that minimises cascade calls subject to
the SLA. The omniscient bound is the smallest cascade cost any
calibration could possibly achieve at this proxy and this target;
it is a within-proxy analogue of the \methBERLB lower bound,
restricted to the calibration.}

{The four ablated calibrations split cleanly along the
deployable / non-deployable axis and along the SLA-safe /
SLA-violating axis.
\emph{(i)~The omniscient bound sets the floor at $297.2$\,s.}
This is the smallest cascade cost any calibration could possibly
achieve on this proxy at $\alpha=0.9$. Any deployable calibration
must sit above it.
\emph{(ii)~The naive empirical calibration dips below the floor at
$295.3$\,s, but only by violating the SLA.} It meets the target on
only $12/17$ queries and incurs the largest violation magnitude
($0.0602$) in the table. The $295.3\!<\!297.2$ gap is exactly the
cost of those violations: the calibration saves cascade calls on
queries it should have cascaded.
\emph{(iii)~\methScaleDoc{}'s smoothed band pays the opposite
cost.} It meets the SLA on all $17/17$ queries but at $363.2$\,s,
$\sim$$22\%$ above the floor, because its uniform smoothing
inflates the per-bin error estimate on \emph{every} score range --
including the well-covered ones whose empirical rate is already
reliable -- and the threshold cascades more documents than the
target actually requires (cf.\ \S\ref{sec:bercp:why_tighter}).
\emph{(iv)~Our calibration sits within $\sim$$10\%$ of the floor.}
$328.7$\,s ($+10.6\%$ over $297.2$) with $16/17$ SLA hits and
violation magnitude $0.0071$: one missed query by a hair. 
The single-knob per-bin blend
(\S\ref{sec:bercp:perbin}) is the only deployable calibration in
the table that sits close to the within-proxy floor on \emph{both
axes simultaneously} -- E2E cost (column 1) and SLA hit count
(column 4).}

\begin{table}[t]
\centering\small
\setlength{\tabcolsep}{4pt}\renewcommand{\arraystretch}{1.18}
\begin{tabular}{@{}llrr@{}}
\toprule
\textbf{Architecture} & \textbf{Training} & E2E (s) & acc$\ge\!0.9$ \\
\midrule
\multicolumn{4}{l}{\emph{Architecture sweep (training fixed at ours)}}\\
\textbf{CE+CB+hybrid (ours)} & \textbf{soft-BCE + PD + cov} & \textbf{328.7} & \textbf{16/17} \\ % correct
Bi-encoder (\methScaleDoc)   & soft-BCE                     & 415.4 & 13/17 \\ % correct
\midrule
\multicolumn{4}{l}{\emph{Training loss for CE/CB}}\\
CE+CB+hybrid & contrastive + PD + cov & 329.6 & 14/17 \\ % correct
CE+CB+hybrid & hard-BCE    + PD + cov & 355.8 & 12/17 \\ % correct
\midrule
\multicolumn{4}{l}{\emph{Training loss for Hybrid}}\\
CE+CB+hybrid & soft-BCE    + PD       & 406.4 & 15/17 \\ % correct
CE+CB+hybrid & soft-BCE    + cov      & 340.5 & 13/17 \\ % correct
%CE+CB+hybrid & soft-BCE               & 329.9 & 15/17 \\ % correct
\midrule
\multicolumn{4}{l}{\emph{Baseline reference}}\\ 
Bi-encoder (\methDefault) & contrastive       & 543.1 & 13/17 \\ % correct 
\bottomrule
\end{tabular}
\caption{{\textbf{Proxy ablation on \dsPubMed.} The $3$
Phase-1-only queries are removed, so all rows are reported on the
same $17$ queries; all proxies are trained on the same Phase-1
labeled documents. PD: primal--dual update on $\Cset$; cov:
coverage regularizer on $\Tset$. Calibration is fixed to our full
per-bin Clopper--Pearson blend throughout.}}
\label{tab:proxy_ablation}
\end{table}

\begin{table}[t]
\centering\small
\setlength{\tabcolsep}{4pt}\renewcommand{\arraystretch}{1.18}
\begin{tabular}{@{}lrrrrr@{}}
\toprule
\textbf{Calibration}                & E2E (s)       & mean acc & min acc & acc$\ge\!0.9$ & SLA viol. \\
\midrule
\textbf{Ours}                       & \textbf{328.7} & \textbf{0.919} & \textbf{0.893} & \textbf{16/17} & \textbf{0.0071} \\
%$-$ BER                             & 330.9 & 0.912 & 0.886 & 13/16 & 0.0063 \\
%$-$ BER $-$ align                   & 331.7 & 0.912 & 0.890 & 13/16 & 0.0063 \\
\methScaleDoc                       & 363.2 & 0.931 & 0.916 & 17/17 & 0.0000 \\
{Naive empirical} & 295.3 & 0.910 & 0.876 & 12/17 & 0.0602 \\
Omniscient bound                    & 297.2 & 0.910 & 0.900 & 17/17 & 0.0000 \\
\bottomrule
\end{tabular}
\caption{
{\textbf{Calibration ablation on \dsPubMed.} Proxy
held fixed at our CE+CB+hybrid with soft-BCE, primal--dual,
and coverage regularizer.
\emph{Naive empirical}: bare per-bin error rate on $\Cset$, 
no safety margin.
\emph{\methScaleDoc}: $64$-bin histogram band with
smoothed yes/no counts.
\emph{Ours}: single-knob per-bin Clopper--Pearson blend of
\S\ref{sec:bercp:perbin}.
\emph{Omniscient bound}: non-deployable; assumes every document's
true label is known and picks the threshold that minimizes
cascade calls subject to the SLA.}}
\label{tab:cal_ablation}
\end{table}

\section{Conclusion}\label{sec:conclusion}

{This paper optimizes the semantic filtering operator, unifies existing cascade plans into a single six-step framework with
four design knobs, and shows that prior cascades have been
optimizing the wrong knobs. 
Within the framework, we propose a token-aware hybrid
proxy, trained for the first time with the oracle's per-document
confidence as a soft label, so the proxy's probability tracks the
oracle's at the deployment target; a single-knob blend of the per-bin
empirical rate with a Clopper--Pearson upper bound, increasing the safety margin \emph{only} on the sparse score ranges that actually need it; an adaptive two-phase model-free and online-proxy composition that
reuses the Phase-1 labels as the Phase-2 training set to reduce the transition cost; and the first use of the oracle's own
per-document Bayes Error Rate as both a per-query difficulty
compass and a non-deployable lower bound on the cost any
proxy-based cascade can achieve. 
The combined plan is $1.6$--$2.0\times$ faster than the previous
deployable state of the art on every corpus and within
$\sim$$10\%$ of a non-deployable omniscient calibration that knows
every label in advance. Further remaining $\sim$$4$--$20\times$
gap to the strict performance upper bound could be achieved with a better proxy and probably through adaptive per-query labeling budgets for training and calibration.}

%and extensions to join, group-by, and
%top-$k$ -- is the natural next step.}

% ============================================================
%  References
% ============================================================
\bibliographystyle{ACM-Reference-Format}
\bibliography{references}

@inproceedings{lee2025nvembed,
  author       = {Chankyu Lee and
                  Rajarshi Roy and
                  Mengyao Xu and
                  Jonathan Raiman and
                  Mohammad Shoeybi and
                  Bryan Catanzaro and
                  Wei Ping},
  title        = {NV-Embed: Improved Techniques for Training LLMs as Generalist Embedding
                  Models},
  booktitle    = {The Thirteenth International Conference on Learning Representations,
                  {ICLR} 2025, Singapore, April 24-28, 2025},
  publisher    = {OpenReview.net},
  year         = {2025},
  url          = {https://openreview.net/forum?id=lgsyLSsDRe},
  bibsource    = {dblp computer science bibliography, https://dblp.org}
}

@inproceedings{devlin2019bert,
  author       = {Jacob Devlin and
                  Ming{-}Wei Chang and
                  Kenton Lee and
                  Kristina Toutanova},
  editor       = {Jill Burstein and
                  Christy Doran and
                  Thamar Solorio},
  title        = {{BERT:} Pre-training of Deep Bidirectional Transformers for Language
                  Understanding},
  booktitle    = {Proceedings of the 2019 Conference of the North American Chapter of
                  the Association for Computational Linguistics: Human Language Technologies,
                  {NAACL-HLT} 2019, Minneapolis, MN, USA, June 2-7, 2019, Volume 1 (Long
                  and Short Papers)},
  pages        = {4171--4186},
  publisher    = {Association for Computational Linguistics},
  year         = {2019},
  url          = {https://doi.org/10.18653/v1/n19-1423},
  doi          = {10.18653/V1/N19-1423},
  bibsource    = {dblp computer science bibliography, https://dblp.org}
}

@article{clopper1934use,
  title={The use of confidence or fiducial limits illustrated in the case of the binomial},
  author={Clopper, Charles J and Pearson, Egon S},
  journal={Biometrika},
  volume={26},
  number={4},
  pages={404--413},
  year={1934},
  publisher={JSTOR}
}

@article{llama3,
  author       = {Llama Team},
  title        = {The Llama 3 Herd of Models},
  journal      = {CoRR},
  volume       = {abs/2407.21783},
  year         = {2024},
  url          = {https://doi.org/10.48550/arXiv.2407.21783},
  doi          = {10.48550/ARXIV.2407.21783},
  eprinttype   = {arXiv},
  eprint       = {2407.21783},
  bibsource    = {dblp computer science bibliography, https://dblp.org}
}

@article{scaledoc,
  author       = {Hengrui Zhang and
                  Yulong Hui and
                  Yihao Liu and
                  Huanchen Zhang},
  title        = {ScaleDoc: Scaling LLM-based Predicates over Large Document Collections},
  journal      = {CoRR},
  volume       = {abs/2509.12610},
  year         = {2025},
  url          = {https://doi.org/10.48550/arXiv.2509.12610},
  doi          = {10.48550/ARXIV.2509.12610},
  eprinttype   = {arXiv},
  eprint       = {2509.12610},
  bibsource    = {dblp computer science bibliography, https://dblp.org}
}

@inproceedings{khattab2020colbert,
  author       = {Omar Khattab and
                  Matei Zaharia},
  editor       = {Jimmy X. Huang and
                  Yi Chang and
                  Xueqi Cheng and
                  Jaap Kamps and
                  Vanessa Murdock and
                  Ji{-}Rong Wen and
                  Yiqun Liu},
  title        = {ColBERT: Efficient and Effective Passage Search via Contextualized
                  Late Interaction over {BERT}},
  booktitle    = {Proceedings of the 43rd International {ACM} {SIGIR} conference on
                  research and development in Information Retrieval, {SIGIR} 2020, Virtual
                  Event, China, July 25-30, 2020},
  pages        = {39--48},
  publisher    = {{ACM}},
  year         = {2020},
  url          = {https://doi.org/10.1145/3397271.3401075},
  doi          = {10.1145/3397271.3401075},
  bibsource    = {dblp computer science bibliography, https://dblp.org}
}

@article{nogueira2019passage,
  author       = {Rodrigo Nogueira and
                  Kyunghyun Cho},
  title        = {Passage Re-ranking with {BERT}},
  journal      = {CoRR},
  volume       = {abs/1901.04085},
  year         = {2019},
  url          = {http://arxiv.org/abs/1901.04085},
  eprinttype   = {arXiv},
  eprint       = {1901.04085},
  bibsource    = {dblp computer science bibliography, https://dblp.org}
}

@book{vovk2005algorithmic,
  title={Algorithmic learning in a random world},
  author={Vovk, Vladimir and Gammerman, Alexander and Shafer, Glenn},
  year={2005},
  publisher={Springer}
}

@article{angelopoulos2021gentle,
  author       = {Anastasios N. Angelopoulos and
                  Stephen Bates},
  title        = {A Gentle Introduction to Conformal Prediction and Distribution-Free
                  Uncertainty Quantification},
  journal      = {CoRR},
  volume       = {abs/2107.07511},
  year         = {2021},
  url          = {https://arxiv.org/abs/2107.07511},
  eprinttype   = {arXiv},
  eprint       = {2107.07511},
  bibsource    = {dblp computer science bibliography, https://dblp.org}
}

@inproceedings{gibbs2021adaptive,
  author       = {Isaac Gibbs and
                  Emmanuel J. Cand{\`{e}}s},
  editor       = {Marc'Aurelio Ranzato and
                  Alina Beygelzimer and
                  Yann N. Dauphin and
                  Percy Liang and
                  Jennifer Wortman Vaughan},
  title        = {Adaptive Conformal Inference Under Distribution Shift},
  booktitle    = {Advances in Neural Information Processing Systems 34: Annual Conference
                  on Neural Information Processing Systems 2021, NeurIPS 2021, December
                  6-14, 2021, virtual},
  pages        = {1660--1672},
  year         = {2021},
  url          = {https://proceedings.neurips.cc/paper/2021/hash/0d441de75945e5acbc865406fc9a2559-Abstract.html},
  bibsource    = {dblp computer science bibliography, https://dblp.org}
}

@inproceedings{zaffran2022adaptive,
  author       = {Margaux Zaffran and
                  Olivier F{\'{e}}ron and
                  Yannig Goude and
                  Julie Josse and
                  Aymeric Dieuleveut},
  editor       = {Kamalika Chaudhuri and
                  Stefanie Jegelka and
                  Le Song and
                  Csaba Szepesv{\'{a}}ri and
                  Gang Niu and
                  Sivan Sabato},
  title        = {Adaptive Conformal Predictions for Time Series},
  booktitle    = {International Conference on Machine Learning, {ICML} 2022, 17-23 July
                  2022, Baltimore, Maryland, {USA}},
  series       = {Proceedings of Machine Learning Research},
  volume       = {162},
  pages        = {25834--25866},
  publisher    = {{PMLR}},
  year         = {2022},
  url          = {https://proceedings.mlr.press/v162/zaffran22a.html},
  bibsource    = {dblp computer science bibliography, https://dblp.org}
}

@article{bates2021distribution,
  author       = {Stephen Bates and
                  Anastasios Angelopoulos and
                  Lihua Lei and
                  Jitendra Malik and
                  Michael I. Jordan},
  title        = {Distribution-free, Risk-controlling Prediction Sets},
  journal      = {J. {ACM}},
  volume       = {68},
  number       = {6},
  pages        = {43:1--43:34},
  year         = {2021},
  url          = {https://doi.org/10.1145/3478535},
  doi          = {10.1145/3478535},
  bibsource    = {dblp computer science bibliography, https://dblp.org}
}

@article{chen2023frugalgpt,
  author       = {Lingjiao Chen and
                  Matei Zaharia and
                  James Zou},
  title        = {FrugalGPT: How to Use Large Language Models While Reducing Cost and
                  Improving Performance},
  journal      = {Trans. Mach. Learn. Res.},
  volume       = {2024},
  year         = {2024},
  url          = {https://openreview.net/forum?id=cSimKw5p6R},
  bibsource    = {dblp computer science bibliography, https://dblp.org}
}

@inproceedings{lu2023probabilistic,
  author       = {Yao Lu and
                  Aakanksha Chowdhery and
                  Srikanth Kandula and
                  Surajit Chaudhuri},
  editor       = {Gautam Das and
                  Christopher M. Jermaine and
                  Philip A. Bernstein},
  title        = {Accelerating Machine Learning Inference with Probabilistic Predicates},
  booktitle    = {Proceedings of the 2018 International Conference on Management of
                  Data, {SIGMOD} Conference 2018, Houston, TX, USA, June 10-15, 2018},
  pages        = {1493--1508},
  publisher    = {{ACM}},
  year         = {2018},
  url          = {https://doi.org/10.1145/3183713.3183751},
  doi          = {10.1145/3183713.3183751},
  bibsource    = {dblp computer science bibliography, https://dblp.org}
}

@article{wang2023zero,
  author       = {Xiang Wei and
                  Xingyu Cui and
                  Ning Cheng and
                  Xiaobin Wang and
                  Xin Zhang and
                  Shen Huang and
                  Pengjun Xie and
                  Jinan Xu and
                  Yufeng Chen and
                  Meishan Zhang and
                  Yong Jiang and
                  Wenjuan Han},
  title        = {Zero-Shot Information Extraction via Chatting with ChatGPT},
  journal      = {CoRR},
  volume       = {abs/2302.10205},
  year         = {2023},
  url          = {https://doi.org/10.48550/arXiv.2302.10205},
  doi          = {10.48550/ARXIV.2302.10205},
  eprinttype   = {arXiv},
  eprint       = {2302.10205},
  bibsource    = {dblp computer science bibliography, https://dblp.org}
}

@inproceedings{wang2011cascade,
  title={Cascade ranking for operational e-commerce search},
  author={Liu, Shichen and Xiao, Fei and Ou, Wenwu and Si, Luo},
  booktitle={Proceedings of the 23rd ACM SIGKDD International Conference on Knowledge Discovery and Data Mining},
  pages={1557--1565},
  year={2017}
}

@inproceedings{bolukbasi2017adaptive,
  author       = {Tolga Bolukbasi and
                  Joseph Wang and
                  Ofer Dekel and
                  Venkatesh Saligrama},
  editor       = {Doina Precup and
                  Yee Whye Teh},
  title        = {Adaptive Neural Networks for Efficient Inference},
  booktitle    = {Proceedings of the 34th International Conference on Machine Learning,
                  {ICML} 2017, Sydney, NSW, Australia, 6-11 August 2017},
  series       = {Proceedings of Machine Learning Research},
  volume       = {70},
  pages        = {527--536},
  publisher    = {{PMLR}},
  year         = {2017},
  url          = {http://proceedings.mlr.press/v70/bolukbasi17a.html},
  bibsource    = {dblp computer science bibliography, https://dblp.org}
}

@book{fukunaga1990introduction,
  title={Introduction to statistical pattern recognition},
  author={Fukunaga, Keinosuke},
  year={2013},
  publisher={Elsevier}
}

@article{northcutt2021confident,
  author       = {Curtis G. Northcutt and
                  Lu Jiang and
                  Isaac L. Chuang},
  title        = {Confident Learning: Estimating Uncertainty in Dataset Labels},
  journal      = {J. Artif. Intell. Res.},
  volume       = {70},
  pages        = {1373--1411},
  year         = {2021},
  url          = {https://doi.org/10.1613/jair.1.12125},
  doi          = {10.1613/JAIR.1.12125},
  bibsource    = {dblp computer science bibliography, https://dblp.org}
}

@article{csv2026,
  title={Beyond Linear LLM Invocation: An Efficient and Effective Semantic Filter Paradigm},
  author={Hou, Nan and Zhao, Kangfei and Xie, Jiadong and Yu, Jeffrey Xu},
  journal={arXiv preprint arXiv:2603.04799},
  year={2026}
}

@article{bargain2025,
  title={Cut Costs, Not Accuracy: LLM-Powered Data Processing with Guarantees},
  author={Zeighami, Sepanta and Shankar, Shreya and Parameswaran, Aditya},
  journal={Proceedings of the ACM on Management of Data},
  volume={3},
  number={6},
  pages={1--26},
  year={2025},
  publisher={ACM New York, NY, USA}
}

@inproceedings{liu2024palimpzest,
  title={Palimpzest: Optimizing ai-powered analytics with declarative query processing},
  author={Liu, Chunwei and Russo, Matthew and Cafarella, Michael and Cao, Lei and Chen, Peter Baile and Chen, Zui and Franklin, Michael and Kraska, Tim and Madden, Samuel and Shahout, Rana and others},
  booktitle={Proceedings of the Conference on Innovative Database Research (CIDR)},
  pages={2},
  year={2025}
}

@article{patel2024lotus,
  author       = {Liana Patel and
                  Siddharth Jha and
                  Carlos Guestrin and
                  Matei Zaharia},
  title        = {{LOTUS:} Enabling Semantic Queries with LLMs Over Tables of Unstructured
                  and Structured Data},
  journal      = {CoRR},
  volume       = {abs/2407.11418},
  year         = {2024},
  url          = {https://doi.org/10.48550/arXiv.2407.11418},
  doi          = {10.48550/ARXIV.2407.11418},
  eprinttype   = {arXiv},
  eprint       = {2407.11418},
  bibsource    = {dblp computer science bibliography, https://dblp.org}
}

@article{kang2020supg,
  author       = {Daniel Kang and
                  Edward Gan and
                  Peter Bailis and
                  Tatsunori Hashimoto and
                  Matei Zaharia},
  title        = {Approximate Selection with Guarantees using Proxies},
  journal      = {Proc. {VLDB} Endow.},
  volume       = {13},
  number       = {11},
  pages        = {1990--2003},
  year         = {2020},
  url          = {http://www.vldb.org/pvldb/vol13/p1990-kang.pdf},
  bibsource    = {dblp computer science bibliography, https://dblp.org}
}

@article{hinton2015distilling,
  author       = {Geoffrey E. Hinton and
                  Oriol Vinyals and
                  Jeffrey Dean},
  title        = {Distilling the Knowledge in a Neural Network},
  journal      = {CoRR},
  volume       = {abs/1503.02531},
  year         = {2015},
  url          = {http://arxiv.org/abs/1503.02531},
  eprinttype   = {arXiv},
  eprint       = {1503.02531},
  bibsource    = {dblp computer science bibliography, https://dblp.org}
}

@inproceedings{sioulas2021speculation,
  author       = {Panagiotis Sioulas and
                  Viktor Sanca and
                  Ioannis Mytilinis and
                  Anastasia Ailamaki},
  title        = {Accelerating Complex Analytics using Speculation},
  booktitle    = {11th Conference on Innovative Data Systems Research, {CIDR} 2021,
                  Virtual Event, January 11-15, 2021, Online Proceedings},
  publisher    = {www.cidrdb.org},
  year         = {2021},
  url          = {https://vldb.org/cidrdb/2021/accelerating-complex-analytics-using-speculation.html},
  bibsource    = {dblp computer science bibliography, https://dblp.org}
}

@inproceedings{geifman,
  author       = {Yonatan Geifman and
                  Ran El{-}Yaniv},
  editor       = {Isabelle Guyon and
                  Ulrike von Luxburg and
                  Samy Bengio and
                  Hanna M. Wallach and
                  Rob Fergus and
                  S. V. N. Vishwanathan and
                  Roman Garnett},
  title        = {Selective Classification for Deep Neural Networks},
  booktitle    = {Advances in Neural Information Processing Systems 30: Annual Conference
                  on Neural Information Processing Systems 2017, December 4-9, 2017,
                  Long Beach, CA, {USA}},
  pages        = {4878--4887},
  year         = {2017},
  url          = {https://proceedings.neurips.cc/paper/2017/hash/4a8423d5e91fda00bb7e46540e2b0cf1-Abstract.html},
  bibsource    = {dblp computer science bibliography, https://dblp.org}
}

@article{shankar2024docetl,
  author       = {Shreya Shankar and
                  Tristan Chambers and
                  Tarak Shah and
                  Aditya G. Parameswaran and
                  Eugene Wu},
  title        = {DocETL: Agentic Query Rewriting and Evaluation for Complex Document
                  Processing},
  journal      = {Proc. {VLDB} Endow.},
  volume       = {18},
  number       = {9},
  pages        = {3035--3048},
  year         = {2025},
  url          = {https://www.vldb.org/pvldb/vol18/p3035-shankar.pdf},
  doi          = {10.14778/3746405.3746426},
  bibsource    = {dblp computer science bibliography, https://dblp.org}
}

@article{chung2026proxy,
  author       = {Yeounoh Chung and
                  Rushabh Desai and
                  Jian He and
                  Yu Xiao and
                  Thibaud Hottelier and
                  Yves{-}Laurent Kom Samo and
                  Pushkar Khadilkar and
                  Xianshun Chen and
                  Sam Idicula and
                  Fatma {\"{O}}zcan and
                  Alon Y. Halevy and
                  Yannis Papakonstantinou},
  title        = {100x Cost {\&} Latency Reduction: Performance Analysis of {AI}
                  Query Approximation using Lightweight Proxy Models},
  journal      = {CoRR},
  volume       = {abs/2603.15970},
  year         = {2026},
  url          = {https://doi.org/10.48550/arXiv.2603.15970},
  doi          = {10.48550/ARXIV.2603.15970},
  eprinttype   = {arXiv},
  eprint       = {2603.15970},
  bibsource    = {dblp computer science bibliography, https://dblp.org}
}

@article{mang2026plop,
  author       = {Qiuyang Mang and
                  Yufan Xiang and
                  Hangrui Zhou and
                  Runyuan He and
                  Jiaxiang Yu and
                  Hanchen Li and
                  Aditya G. Parameswaran and
                  Alvin Cheung},
  title        = {{PLOP:} Cost-Based Placement of Semantic Operators in Hybrid Query
                  Plans},
  journal      = {CoRR},
  volume       = {abs/2604.09944},
  year         = {2026},
  url          = {https://doi.org/10.48550/arXiv.2604.09944},
  doi          = {10.48550/ARXIV.2604.09944},
  eprinttype   = {arXiv},
  eprint       = {2604.09944},
  bibsource    = {dblp computer science bibliography, https://dblp.org}
}

@article{kossmann2026ken,
  author       = {Ferdinand Kossmann and
                  Ziniu Wu and
                  Alex Turk and
                  Nesime Tatbul and
                  Lei Cao and
                  Samuel Madden},
  title        = {Ken: An Execution Engine for Unstructured Database Systems},
  journal      = {Proc. {VLDB} Endow.},
  volume       = {19},
  number       = {5},
  pages        = {902--916},
  year         = {2026},
  url          = {https://www.vldb.org/pvldb/vol19/p902-kossmann.pdf},
  bibsource    = {dblp computer science bibliography, https://dblp.org}
}

@book{boyd2004convex,
  title={Convex optimization},
  author={Boyd, Stephen and Vandenberghe, Lieven},
  year={2004},
  publisher={Cambridge university press}
}

@inproceedings{reimers2019sentence,
  author       = {Nils Reimers and
                  Iryna Gurevych},
  editor       = {Kentaro Inui and
                  Jing Jiang and
                  Vincent Ng and
                  Xiaojun Wan},
  title        = {Sentence-BERT: Sentence Embeddings using Siamese BERT-Networks},
  booktitle    = {Proceedings of the 2019 Conference on Empirical Methods in Natural
                  Language Processing and the 9th International Joint Conference on
                  Natural Language Processing, {EMNLP-IJCNLP} 2019, Hong Kong, China,
                  November 3-7, 2019},
  pages        = {3980--3990},
  publisher    = {Association for Computational Linguistics},
  year         = {2019},
  url          = {https://doi.org/10.18653/v1/D19-1410},
  doi          = {10.18653/V1/D19-1410},
  bibsource    = {dblp computer science bibliography, https://dblp.org}
}

@inproceedings{karpukhin2020dense,
  author       = {Vladimir Karpukhin and
                  Barlas Oguz and
                  Sewon Min and
                  Patrick Lewis and
                  Ledell Wu and
                  Sergey Edunov and
                  Danqi Chen and
                  Wen{-}tau Yih},
  editor       = {Bonnie Webber and
                  Trevor Cohn and
                  Yulan He and
                  Yang Liu},
  title        = {Dense Passage Retrieval for Open-Domain Question Answering},
  booktitle    = {Proceedings of the 2020 Conference on Empirical Methods in Natural
                  Language Processing, {EMNLP} 2020, Online, November 16-20, 2020},
  pages        = {6769--6781},
  publisher    = {Association for Computational Linguistics},
  year         = {2020},
  url          = {https://doi.org/10.18653/v1/2020.emnlp-main.550},
  doi          = {10.18653/V1/2020.EMNLP-MAIN.550},
  bibsource    = {dblp computer science bibliography, https://dblp.org}
}

@inproceedings{santhanam2022colbertv2,
  author       = {Keshav Santhanam and
                  Omar Khattab and
                  Jon Saad{-}Falcon and
                  Christopher Potts and
                  Matei Zaharia},
  editor       = {Marine Carpuat and
                  Marie{-}Catherine de Marneffe and
                  Iv{\'{a}}n Vladimir Meza Ru{\'{\i}}z},
  title        = {ColBERTv2: Effective and Efficient Retrieval via Lightweight Late
                  Interaction},
  booktitle    = {Proceedings of the 2022 Conference of the North American Chapter of
                  the Association for Computational Linguistics: Human Language Technologies,
                  {NAACL} 2022, Seattle, WA, United States, July 10-15, 2022},
  pages        = {3715--3734},
  publisher    = {Association for Computational Linguistics},
  year         = {2022},
  url          = {https://doi.org/10.18653/v1/2022.naacl-main.272},
  doi          = {10.18653/V1/2022.NAACL-MAIN.272},
  bibsource    = {dblp computer science bibliography, https://dblp.org}
}

@inproceedings{lin2021pretrained,
  author       = {Andrew Yates and
                  Rodrigo Nogueira and
                  Jimmy Lin},
  editor       = {Fernando Diaz and
                  Chirag Shah and
                  Torsten Suel and
                  Pablo Castells and
                  Rosie Jones and
                  Tetsuya Sakai},
  title        = {Pretrained Transformers for Text Ranking: {BERT} and Beyond},
  booktitle    = {{SIGIR} '21: The 44th International {ACM} {SIGIR} Conference on Research
                  and Development in Information Retrieval, Virtual Event, Canada, July
                  11-15, 2021},
  pages        = {2666--2668},
  publisher    = {{ACM}},
  year         = {2021},
  url          = {https://doi.org/10.1145/3404835.3462812},
  doi          = {10.1145/3404835.3462812},
  bibsource    = {dblp computer science bibliography, https://dblp.org}
}

@inproceedings{bajaj2016msmarco,
  author       = {Tri Nguyen and
                  Mir Rosenberg and
                  Xia Song and
                  Jianfeng Gao and
                  Saurabh Tiwary and
                  Rangan Majumder and
                  Li Deng},
  editor       = {Tarek Richard Besold and
                  Antoine Bordes and
                  Artur S. d'Avila Garcez and
                  Greg Wayne},
  title        = {{MS} {MARCO:} {A} Human Generated MAchine Reading COmprehension Dataset},
  booktitle    = {Proceedings of the Workshop on Cognitive Computation: Integrating
                  neural and symbolic approaches 2016 co-located with the 30th Annual
                  Conference on Neural Information Processing Systems {(NIPS} 2016),
                  Barcelona, Spain, December 9, 2016},
  series       = {{CEUR} Workshop Proceedings},
  volume       = {1773},
  publisher    = {CEUR-WS.org},
  year         = {2016},
  url          = {https://ceur-ws.org/Vol-1773/CoCoNIPS\_2016\_paper9.pdf},
  bibsource    = {dblp computer science bibliography, https://dblp.org}
}

@article{craswell2020trec,
  author       = {Nick Craswell and
                  Bhaskar Mitra and
                  Emine Yilmaz and
                  Daniel Campos and
                  Ellen M. Voorhees},
  title        = {Overview of the {TREC} 2019 deep learning track},
  journal      = {CoRR},
  volume       = {abs/2003.07820},
  year         = {2020},
  url          = {https://arxiv.org/abs/2003.07820},
  eprinttype   = {arXiv},
  eprint       = {2003.07820},
  bibsource    = {dblp computer science bibliography, https://dblp.org}
}

@inproceedings{thakur2021beir,
  author       = {Nandan Thakur and
                  Nils Reimers and
                  Andreas R{\"{u}}ckl{\'{e}} and
                  Abhishek Srivastava and
                  Iryna Gurevych},
  editor       = {Joaquin Vanschoren and
                  Sai{-}Kit Yeung},
  title        = {{BEIR:} {A} Heterogeneous Benchmark for Zero-shot Evaluation of Information
                  Retrieval Models},
  booktitle    = {Proceedings of the Neural Information Processing Systems Track on
                  Datasets and Benchmarks 1, NeurIPS Datasets and Benchmarks 2021, December
                  2021, virtual},
  year         = {2021},
  url          = {https://datasets-benchmarks-proceedings.neurips.cc/paper/2021/hash/65b9eea6e1cc6bb9f0cd2a47751a186f-Abstract-round2.html},
  bibsource    = {dblp computer science bibliography, https://dblp.org}
}

@inproceedings{muennighoff2023mteb,
  author       = {Niklas Muennighoff and
                  Nouamane Tazi and
                  Lo{\"{\i}}c Magne and
                  Nils Reimers},
  editor       = {Andreas Vlachos and
                  Isabelle Augenstein},
  title        = {{MTEB:} Massive Text Embedding Benchmark},
  booktitle    = {Proceedings of the 17th Conference of the European Chapter of the
                  Association for Computational Linguistics, {EACL} 2023, Dubrovnik,
                  Croatia, May 2-6, 2023},
  pages        = {2006--2029},
  publisher    = {Association for Computational Linguistics},
  year         = {2023},
  url          = {https://doi.org/10.18653/v1/2023.eacl-main.148},
  doi          = {10.18653/V1/2023.EACL-MAIN.148},
  bibsource    = {dblp computer science bibliography, https://dblp.org}
}

@article{wang2022e5,
  author       = {Liang Wang and
                  Nan Yang and
                  Xiaolong Huang and
                  Binxing Jiao and
                  Linjun Yang and
                  Daxin Jiang and
                  Rangan Majumder and
                  Furu Wei},
  title        = {Text Embeddings by Weakly-Supervised Contrastive Pre-training},
  journal      = {CoRR},
  volume       = {abs/2212.03533},
  year         = {2022},
  url          = {https://doi.org/10.48550/arXiv.2212.03533},
  doi          = {10.48550/ARXIV.2212.03533},
  eprinttype   = {arXiv},
  eprint       = {2212.03533},
  bibsource    = {dblp computer science bibliography, https://dblp.org}
}

@article{touvron2023llama,
  author       = {Hugo Touvron and
                  Thibaut Lavril and
                  Gautier Izacard and
                  Xavier Martinet and
                  Marie{-}Anne Lachaux and
                  Timoth{\'{e}}e Lacroix and
                  Baptiste Rozi{\`{e}}re and
                  Naman Goyal and
                  Eric Hambro and
                  Faisal Azhar and
                  Aur{\'{e}}lien Rodriguez and
                  Armand Joulin and
                  Edouard Grave and
                  Guillaume Lample},
  title        = {LLaMA: Open and Efficient Foundation Language Models},
  journal      = {CoRR},
  volume       = {abs/2302.13971},
  year         = {2023},
  url          = {https://doi.org/10.48550/arXiv.2302.13971},
  doi          = {10.48550/ARXIV.2302.13971},
  eprinttype   = {arXiv},
  eprint       = {2302.13971},
  bibsource    = {dblp computer science bibliography, https://dblp.org}
}

@article{bertsekas1999nonlinear,
  title={Nonlinear programming},
  author={Bertsekas, Dimitri P},
  journal={Journal of the Operational Research Society},
  volume={48},
  number={3},
  pages={334--334},
  year={1997},
  publisher={Taylor \& Francis}
}

@article{cotter2019two,
  author       = {Andrew Cotter and
                  Heinrich Jiang and
                  Maya R. Gupta and
                  Serena Lutong Wang and
                  Taman Narayan and
                  Seungil You and
                  Karthik Sridharan},
  title        = {Optimization with Non-Differentiable Constraints with Applications
                  to Fairness, Recall, Churn, and Other Goals},
  journal      = {J. Mach. Learn. Res.},
  volume       = {20},
  pages        = {172:1--172:59},
  year         = {2019},
  url          = {https://jmlr.org/papers/v20/18-616.html},
  bibsource    = {dblp computer science bibliography, https://dblp.org}
}

@article{brown2001interval,
  title={Interval estimation for a binomial proportion},
  author={Brown, Lawrence D and Cai, T Tony and DasGupta, Anirban},
  journal={Statistical science},
  volume={16},
  number={2},
  pages={101--133},
  year={2001},
  publisher={Institute of Mathematical Statistics}
}

@article{wilson1927probable,
  title={Probable inference, the law of succession, and statistical inference},
  author={Wilson, Edwin B},
  journal={Journal of the American Statistical Association},
  volume={22},
  number={158},
  pages={209--212},
  year={1927},
  publisher={Taylor \& Francis}
}

@inproceedings{chen2020simclr,
  author       = {Ting Chen and
                  Simon Kornblith and
                  Mohammad Norouzi and
                  Geoffrey E. Hinton},
  title        = {A Simple Framework for Contrastive Learning of Visual Representations},
  booktitle    = {Proceedings of the 37th International Conference on Machine Learning,
                  {ICML} 2020, 13-18 July 2020, Virtual Event},
  series       = {Proceedings of Machine Learning Research},
  volume       = {119},
  pages        = {1597--1607},
  publisher    = {{PMLR}},
  year         = {2020},
  url          = {http://proceedings.mlr.press/v119/chen20j.html},
  bibsource    = {dblp computer science bibliography, https://dblp.org}
}

@inproceedings{gao2021simcse,
  author       = {Tianyu Gao and
                  Xingcheng Yao and
                  Danqi Chen},
  editor       = {Marie{-}Francine Moens and
                  Xuanjing Huang and
                  Lucia Specia and
                  Scott Wen{-}tau Yih},
  title        = {SimCSE: Simple Contrastive Learning of Sentence Embeddings},
  booktitle    = {Proceedings of the 2021 Conference on Empirical Methods in Natural
                  Language Processing, {EMNLP} 2021, Virtual Event / Punta Cana, Dominican
                  Republic, 7-11 November, 2021},
  pages        = {6894--6910},
  publisher    = {Association for Computational Linguistics},
  year         = {2021},
  url          = {https://doi.org/10.18653/v1/2021.emnlp-main.552},
  doi          = {10.18653/V1/2021.EMNLP-MAIN.552},
  bibsource    = {dblp computer science bibliography, https://dblp.org}
}

@inproceedings{pereyra2017regularizing,
  author       = {Gabriel Pereyra and
                  George Tucker and
                  Jan Chorowski and
                  Lukasz Kaiser and
                  Geoffrey E. Hinton},
  title        = {Regularizing Neural Networks by Penalizing Confident Output Distributions},
  booktitle    = {5th International Conference on Learning Representations, {ICLR} 2017,
                  Toulon, France, April 24-26, 2017, Workshop Track Proceedings},
  publisher    = {OpenReview.net},
  year         = {2017},
  url          = {https://openreview.net/forum?id=HyhbYrGYe},
  bibsource    = {dblp computer science bibliography, https://dblp.org}
}

\end{document}